\documentclass{IEEEoj}
\usepackage{cite}
\usepackage{amsmath,amssymb,amsfonts,bbm}
\usepackage{algorithmic}
\usepackage{graphicx,color}
\usepackage{textcomp}
\def\BibTeX{{\rm B\kern-.05em{\sc i\kern-.025em b}\kern-.08em
    T\kern-.1667em\lower.7ex\hbox{E}\kern-.125emX}}
\AtBeginDocument{\definecolor{ojcolor}{cmyk}{0.93,0.59,0.15,0.02}}

\usepackage{xcolor}         % colors
\definecolor{links}{rgb}{0.0,0,0.9}   % blue
\definecolor{urls}{rgb}{0,0,0.9}    % blue
\definecolor{cites}{rgb}{0.0,0.0,0.9}   % blue
\usepackage[colorlinks,hyperindex,linkcolor=links,citecolor=cites,urlcolor=urls]{hyperref} % generates colored
\ifCLASSINFOpdf
   \usepackage{graphicx}
   \usepackage{caption}
  \graphicspath{{figures/}}
   \DeclareGraphicsExtensions{.pdf,.jpeg,.png}
\else
\fi
\usepackage{url}
\usepackage{tcolorbox}
\usepackage{color,soul}
\usepackage{cancel}
\usepackage[normalem]{ulem}

\usepackage{subcaption}
 \def\authorrefmark#1{\ensuremath{^{\textbf{#1}}}}

 \newcommand {\Tcp} {T_{\rm{cp}}}
 
 \usepackage{glossaries}
 \glsdisablehyper
 \newglossaryentry{simo}
{
  name={SIMO},
  description={single-input multiple-output},
  first={\glsentrydesc{simo} (\glsentrytext{simo})}
}

\newglossaryentry{mmse}
{
  name={MMSE},
  description={minimum mean-square error},
  first={minimum mean-square error (\glsentrytext{mmse})}
}

\newglossaryentry{lmmse}
{
  name={LMMSE},
  description={linear minimum mean-square error},
  first={linear MMSE (\glsentrytext{lmmse})}
}

\newglossaryentry{mf}
{
  name={MF},
  description={matched filtering},
  first={matched filtering (\glsentrytext{mf})}
}

\newglossaryentry{qlmmse}
{
  name={QLMMSE},
  description={quasilinear MMSE},
  first={\glsentrydesc{qlmmse} (\glsentrytext{qlmmse})}
}

\newglossaryentry{na}
{
  name={NA},
  description={normal approximation},
  first={\glsentrydesc{na} (\glsentrytext{na})},
  plural={NAs},
  descriptionplural={normal approximations},
  firstplural={\glsentrydescplural{na} (\glsentryplural{na})}
}

\newglossaryentry{zf}
{
  name={ZF},
  description={zero-forcing},
  first={\glsentrydesc{zf} (\glsentrytext{zf})}
}

\newglossaryentry{rzf}
{
  name={RZF},
  description={regularized zero-forcing},
  first={\glsentrydesc{rzf} (\glsentrytext{rzf})}
}

\newglossaryentry{sir}
{
  name={SIR},
  description={signal-to-interference ratio},
  first={\glsentrydesc{sir} (\glsentrytext{sir})}
}

\newglossaryentry{snr}
{
  name={SNR},
  description={signal-to-noise ratio},
  first={signal-to-noise ratio (\glsentrytext{snr})}
}

\newglossaryentry{mse}
{
  name={MSE},
  description={mean-squared error},
  first={mean-squared error (\glsentrytext{mse})}
}
\newglossaryentry{mmmse}
{
  name={M-MMSE},
  description={multicell MMSE},
  first={\glsentrydesc{mmmse} (\glsentrytext{mmmse})}
}

\newglossaryentry{ls}
{
  name={LS},
  description={least-squares},
  first={\glsentrydesc{ls} (\glsentrytext{ls})}
}

\newglossaryentry{mr}
{
  name={MR},
  description={maximum ratio},
  first={\glsentrydesc{mr} (\glsentrytext{mr})}
}

\newglossaryentry{ue}
{
  name={UE},
  description={user equipment},
  first={\glsentrydesc{ue} (\glsentrytext{ue})},
  plural={UEs},
  descriptionplural={user equipments},
  firstplural={\glsentrydescplural{ue} (\glsentryplural{ue})}
}

\newglossaryentry{ap}
{
  name={AP},
  description={access point},
  first={\glsentrydesc{ap} (\glsentrytext{ap})},
  plural={APs},
  descriptionplural={access points},
  firstplural={\glsentrydescplural{ap} (\glsentryplural{ap})}
}

\newglossaryentry{cpu}
{
  name={CPU},
  description={central processing unit},
  first={\glsentrydesc{cpu} (\glsentrytext{cpu})},
  plural={CPUs},
  descriptionplural={central processing units},
  firstplural={\glsentrydescplural{cpu} (\glsentryplural{cpu})}
}

\newglossaryentry{bs}
{
  name={BS},
  description={base station},
  first={\glsentrydesc{bs} (\glsentrytext{bs})},
  plural={BSs},
  descriptionplural={base stations},
  firstplural={\glsentrydescplural{bs} (\glsentryplural{bs})}
}

\newglossaryentry{ostbc}
{
  name={OSTBC},
  description={orthogonal space-time block code},
  first={\glsentrydesc{ostbc} (\glsentrytext{ostbc})},
  plural={OSTBCs},
  descriptionplural={orthogonal space-time block codes},
  firstplural={\glsentrydescplural{ostbc} (\glsentryplural{ostbc})}
}

\newglossaryentry{biawgn}
{
  name={bi-AWGN},
  description={binary-input additive white Gaussian noise},
  first={\glsentrydesc{biawgn} (\glsentrytext{biawgn})}
}

\newglossaryentry{flnf}
{
  name={FLNF},
  description={fixed-length no-feedback},
  first={\glsentrydesc{flnf} (\glsentrytext{flnf})}
}

\newglossaryentry{crc}
{
  name={CRC},
  description={cyclic redundancy check},
  first={\glsentrydesc{crc} (\glsentrytext{crc})}
}

\newglossaryentry{bsc}
{
  name={BSC},
  description={binary symmetric channel},
  first={\glsentrydesc{bsc} (\glsentrytext{bsc})}
  plural={BSCs},
  descriptionplural={binary symmetric channels},
  firstplural={\glsentrydescplural{bsc} (\glsentryplural{bsc})}
}

\newglossaryentry{bec}
{
  name={BEC},
  description={binary-erasure channel},
  first={\glsentrydesc{bec} (\glsentrytext{bec})}
}
\newglossaryentry{awgn}
{
  name={AWGN},
  description={additive white Gaussian noise},
  first={additive white Gaussian noise (\glsentrytext{awgn})}
}
\newglossaryentry{3gpp}
{
  name={3GPP},
  description={3rd generation partnership project},
  first={\glsentrydesc{3gpp} (\glsentrytext{3gpp})}
}

\newglossaryentry{dl}
{
  name={DL},
  description={downlink},
  first={\glsentrydesc{dl} (\glsentrytext{dl})}
}

\newglossaryentry{LED}
{
  name={LED},
  description={light emitting diode},
  first={\glsentrydesc{LED} (\glsentrytext{LED})},
  plural={LEDs},
  descriptionplural={light emitting diodes},
  firstplural={\glsentrydescplural{LED} (\glsentryplural{LED})}
}

\newglossaryentry{lte}
{
  name={LTE},
  description={long term evolution},
  first={\glsentrydesc{lte} (\glsentrytext{lte})}
}
\newglossaryentry{ofdm}
{
  name={OFDM},
  description={orthogonal frequency-division multiplexing},
  first={orthogonal frequency-division multiplexing (\glsentrytext{ofdm})}
}

\newglossaryentry{ul}
{
  name={UL},
  description={uplink},
  first={\glsentrydesc{ul} (\glsentrytext{ul})}
}
\newglossaryentry{rb}
{
  name={RB},
  description={resource block},
  first={\glsentrydesc{rb} (\glsentrytext{rb})},
  plural={RBs},
  descriptionplural={resource blocks},
  firstplural={\glsentrydescplural{rb} (\glsentryplural{rb})}
}
\newglossaryentry{cu}
{
  name={CU},
  description={channel use},
  first={\glsentrydesc{cu} (\glsentrytext{cu})},
  plural={CUs},
  descriptionplural={channel uses},
  firstplural={\glsentrydescplural{cu} (\glsentryplural{cu})}
}
\newglossaryentry{tti}
{
  name={TTI},
  description={transmission time interval},
  first={\glsentrydesc{tti} (\glsentrytext{tti})},
  plural={TTI's},
  descriptionplural={transmission time intervals},
  firstplural={\glsentrydescplural{tti} (\glsentryplural{tti})}
}
\newglossaryentry{iid}
{
  name={i.i.d.},
  description={independent and identically distributed},
  first={independent and identically distributed (\glsentrytext{iid})}
}
\newglossaryentry{psd}
{
  name={PSD},
  description={power spectral density},
  first={\glsentrydesc{psd} (\glsentrytext{psd})}
}

\newglossaryentry{mimo}
{
  name={MIMO},
  description={multiple-input multiple-output},
  first={multiple-input multiple-output (\glsentrytext{mimo})}
}

\newglossaryentry{bler}
{
  name={BLER},
  description={block error probability},
  first={\glsentrydesc{bler} (\glsentrytext{bler})}
}
\newglossaryentry{mtc}
{
  name={mMTC},
  description={massive machine-type communication},
  first={massive machine-type communication (\glsentrytext{mtc})}
}
\newacronym{csi}{CSI}{channel state information}
\newglossaryentry{dvbt}
{
  name={DVB-T},
  description={terrestrial digital video broadcasting},
  first={\glsentrydesc{dvbt} (\glsentrytext{dvbt})}
}
\newglossaryentry{pat}
{
  name={PAT},
  description={pilot-assisted transmission},
  first={\glsentrydesc{pat} (\glsentrytext{pat})}
}
\newglossaryentry{ml-prob}
{
  name={ML},
  description={maximum likelihood},
  first={\glsentrydesc{ml-prob} (\glsentrytext{ml-prob})}
}
\newglossaryentry{dt}
{
  name={DT},
  description={dependence-testing},
  first={\glsentrydesc{dt} (\glsentrytext{dt})}
}
\newglossaryentry{ustm}
{
  name={USTM},
  description={unitary space-time modulation},
  first={\glsentrydesc{ustm} (\glsentrytext{ustm})}
}
\newglossaryentry{siso}
{
  name={SISO},
  description={single-input single-output},
  first={\glsentrydesc{siso} (\glsentrytext{siso})}
}
\newglossaryentry{snn}
{
  name={SNN},
  description={scaled nearest-neighbor},
  first={\glsentrydesc{snn} (\glsentrytext{snn})}
}

\newglossaryentry{snnd}
{
  name={SNND},
  description={scaled nearest-neighbor decoding},
  first={\glsentrydesc{snnd} (\glsentrytext{snnd})}
}
\newglossaryentry{rcus}
{
  name={RCUs},
  description={random-coding union bound with parameter $s$},
  first={\glsentrydesc{rcus} (\glsentrytext{rcus})}
}
\newglossaryentry{osd}
{
  name={OSD},
  description={ordered statistics decoding},
  first={\glsentrydesc{osd} (\glsentrytext{osd})}
}
\newglossaryentry{llr}
{
  name={LLR},
  description={log-likelihood ratio},
  first={\glsentrydesc{llr} (\glsentrytext{llr})}
   plural={LLR's},
  descriptionplural={log-likelihood ratios},
  firstplural={\glsentrydescplural{llr} (\glsentryplural{llr})}
}
\newglossaryentry{qpsk}
{
  name={QPSK},
  description={quadrature phase shift keying},
  first={quadrature phase shift keying (\glsentrytext{qpsk})}
}
\newglossaryentry{nlos}
{
  name={NLOS},
  description={non-line-of-sight},
  first={\glsentrydesc{nlos} (\glsentrytext{nlos})}
}
\newglossaryentry{los}
{
  name={LOS},
  description={line-of-sight},
  first={\glsentrydesc{los} (\glsentrytext{los})}
}
\newglossaryentry{mgf}
{
  name={MGF},
  description={moment-generating function},
  first={\glsentrydesc{mgf} (\glsentrytext{mgf})}
}
\newglossaryentry{cgf}
{
  name={CGF},
  description={cumulant-generating function},
  first={\glsentrydesc{cgf} (\glsentrytext{cgf})},
  plural={CGFs},
  descriptionplural={cumulant-generating functions},
  firstplural={\glsentrydescplural{cgf} (\glsentryplural{cgf})}
}
\newglossaryentry{pdf}
{
  name={PDF},
  description={probability density function},
  first={probability density function (\glsentrytext{pdf})},
   plural={PDFs},
  descriptionplural={probability density functions},
  firstplural={probability density functions (\glsentryplural{pdf})}
}

\newglossaryentry{ccdf}
{
  name={CCDF},
  description={complementary cumulative distribution function},
  first={\glsentrydesc{ccdf} (\glsentrytext{ccdf})}
}

\newglossaryentry{cdf}
{
  name={CDF},
  description={cumulative distribution function},
  first={\glsentrydesc{cdf} (\glsentrytext{cdf})}
}

\newglossaryentry{gmi}
{
  name={GMI},
  description={generalized mutual information},
  first={\glsentrydesc{gmi} (\glsentrytext{gmi})}
}

\newglossaryentry{arq}
{
  name={ARQ},
  description={automatic repeat request},
  first={\glsentrydesc{arq} (\glsentrytext{arq})}
}

\newglossaryentry{harq}
{
  name={HARQ},
  description={hybrid automatic repeat request},
  first={\glsentrydesc{harq} (\glsentrytext{harq})}
}

\newglossaryentry{fbl}
{
  name={FBL-NF},
  description={fixed blocklength with no feedback},
  first={\glsentrydesc{fbl} (\glsentrytext{fbl})}
}
\newglossaryentry{ssnf}
{
  name={SS-NF},
  description={single-shot no-feedback},
  first={\glsentrydesc{ssnf} (\glsentrytext{ssnf})}
}

\newacronym{urllc}{URLLC}{ultra-reliable low-latency communications}

\newglossaryentry{vlsf}
{
  name={VLSF},
  description={Variable-length stop-feedback},
  first={\glsentrydesc{vlsf} (\glsentrytext{vlsf})}
}

\newglossaryentry{vlnsf}
{
  name={VLNSF},
  description={variable-length noisy stop feedback},
  first={\glsentrydesc{vlnsf} (\glsentrytext{vlnsf})}
}

\newglossaryentry{dmt}
{
  name={DMT},
  description={diversity-multiplexing tradeoff},
  first={\glsentrydesc{dmt} (\glsentrytext{dmt})}
}

\newglossaryentry{dmdt}
{
  name={DMDT},
  description={diversity-multiplexing-delay tradeoff},
  first={\glsentrydesc{dmdt} (\glsentrytext{dmdt})}
}

\newglossaryentry{tddt}
{
  name={TDDT},
  description={throughput-diversity-delay tradeoff},
  first={\glsentrydesc{tddt} (\glsentrytext{tddt})}
}

\newglossaryentry{tdd}
{
  name={TDD},
  description={time-division duplex},
  first={\glsentrydesc{tdd} (\glsentrytext{tdd})}
}

\newglossaryentry{stbc}
{
  name={STBC},
  description={space-time block-codes},
  first={\glsentrydesc{stbc} (\glsentrytext{stbc})},
  plural={STBC's},
  descriptionplural={space-time block-codes},
  firstplural={\glsentrydescplural{stbc} (\glsentryplural{stbc})}
}

\newglossaryentry{harqir}
{
  name={HARQ-IR},
  description={hybrid automatic repeat request with incremental redundancy},
  first={\glsentrydesc{harqir} (\glsentrytext{harqir})}
}

\newglossaryentry{harqcc}
{
  name={HARQ-CC},
  description={hybrid automatic repeat request with chase combining},
  first={\glsentrydesc{harqcc} (\glsentrytext{harqcc})}
}

\newglossaryentry{wp1}
{
  name={w.p.1.},
  description={with probability one},
  first={\glsentrydesc{wp1} (\glsentrytext{wp1})}
}

\newglossaryentry{vlff}
{
  name={VLFF},
  description={variable-length full feedback},
  first={\glsentrydesc{vlff} (\glsentrytext{vlff})}
}

\newglossaryentry{dmc}
{
  name={DMC},
  description={discrete memoryless channel},
  first={\glsentrydesc{dmc} (\glsentrytext{dmc})}
  plural={DMCs},
  descriptionplural={discrete memoryless channels},
  firstplural={\glsentrydescplural{dmc} (\glsentryplural{dmc})}
}

\newglossaryentry{dnn}
{
  name={DNN},
  description={deep neural network},
  first={deep neural network (\glsentrytext{dnn})}
  plural={DNNs},
  descriptionplural={deep neural networks},
  firstplural={deep neural networks (\glsentryplural{dnn})}
}

\newglossaryentry{nn}
{
  name={NN},
  description={neural network},
  first={neural network (\glsentrytext{nn})}
  plural={NNs},
  descriptionplural={neural networks},
  firstplural={neural networks (\glsentryplural{nn})}
}

\newglossaryentry{cnn}
{
  name={CNN},
  description={convolutional neural network},
  first={convolutional neural network (\glsentrytext{cnn})}
  plural={CNNs},
  descriptionplural={convolutional neural networks},
  firstplural={convolutional neural networks (\glsentryplural{cnn})}
}

\newglossaryentry{ber}
{
  name={BER},
  description={bit error rate},
  first={bit error rate (\glsentrytext{ber})}
}

\newglossaryentry{scss}
{
  name={SCSS},
  description={\emph{single-channel} source separation},
  first={\glsentrydesc{scss} (\glsentrytext{scss})}
}

\newglossaryentry{fft}
{
  name={FFT},
  description={fast Fourier transform},
  first={fast Fourier transform (\glsentrytext{fft})}
}

\newglossaryentry{dft}
{
  name={DFT},
  description={discrete Fourier transform},
  first={discrete Fourier transform (\glsentrytext{dft})}
}

\newglossaryentry{idft}
{
  name={IDFT},
  description={inverse discrete Fourier transform},
  first={inverse discrete Fourier transform (\glsentrytext{idft})}
}

\newglossaryentry{cp}
{
  name={CP},
  description={cyclic prefix},
  first={cyclic prefix (\glsentrytext{cp})}
}

\newglossaryentry{soi}
{
  name={SOI},
  description={signal of interest},
  first={signal of interest (\glsentrytext{soi})}
}

\newglossaryentry{map}
{
  name={MAP},
  description={maximum \emph{a posteriori}},
  first={maximum \emph{a posteriori} (\glsentrytext{map})}
}

\newglossaryentry{usc}
{
  name={TDC},
  description={``temporal-diversity" condition},
  first={\glsentrydesc{usc} (\glsentrytext{usc})}
}

\newglossaryentry{qam}
{
  name={QAM},
  description={quadrature amplitude modulation},
  first={quadrature amplitude modulation (\glsentrytext{qam})}
}

\newglossaryentry{mit}
{
  name={MIT},
  description={Massachusetts Institute of Technology},
  first={\glsentrydesc{mit} (\glsentrytext{mit})}
}

\newglossaryentry{rf}
{
  name={RF},
  description={radio-frequency},
  first={radio-frequency (\glsentrytext{rf})}
}

\newglossaryentry{sdr}
{
  name={SDR},
  description={software defined radio},
  first={software defined radio (\glsentrytext{sdr})}
}

\newglossaryentry{ism}
{
  name={ISM},
  description={industrial, scientific, and medical},
  first={industrial, scientific, and medical (\glsentrytext{ism})}
}

\newglossaryentry{ai}
{
  name={AI},
  description={artificial intelligence},
  first={artificial intelligence (\glsentrytext{ai})}
}

\newglossaryentry{ml}
{
  name={ML},
  description={machine learning},
  first={machine learning (\glsentrytext{ml})}
}

\newglossaryentry{srcsp}
{
  name={SRCSP},
  description={source separation},
  first={\glsentrydesc{srcsp} (\glsentrytext{srcsp})}
}

\newglossaryentry{icassp}
{
  name={ICASSP},
  description={International Conference on Acoustics, Speech, \& Signal Processing},
  first={International Conference on Acoustics, Speech, \& Signal Processing (\glsentrytext{icassp})}
}

\newglossaryentry{sos}
{
  name={SOS},
  description={second-order statistics},
  first={second-order statistics (\glsentrytext{sos})}
}

\newglossaryentry{sinr}
{
  name={SINR},
  description={signal-to-interference-and-noise ratio},
  first={signal-to-interference-and-noise ratio (\glsentrytext{sinr})}
}

\newglossaryentry{fsk}
{
  name={FSK},
  description={frequency shift keying},
  first={frequency shift keying (\glsentrytext{fsk})}
}

\newglossaryentry{ask}
{
  name={ASK},
  description={amplitude shift keying},
  first={amplitude shift keying (\glsentrytext{ask})}
}

\newglossaryentry{psk}
{
  name={PSK},
  description={phase shift keying},
  first={phase shift keying (\glsentrytext{qpsk})}
}

\newglossaryentry{css}
{
  name={CSS},
  description={chirp spread spectrum},
  first={chirp spread spectrum (\glsentrytext{css})}
}

 \usepackage{vmr-symbols-vecbold}
 \usepackage{standard-macros}
 
 \newcommand{\lmmse}{\text{\tiny LMMSE}}
 \newcommand{\mf}{\text{\tiny MF}}

\makeatletter
\newcommand\footnoteref[1]{\protected@xdef\@thefnmark{\ref{#1}}\@footnotemark}

\makeatletter

\makeatletter
\def\@seccntformat#1{%
  \csname the#1\endcsname.\,\, % Inserts the section number (e.g., IV-A. or IV-A.1)) followed by a space
}
\makeatother
\makeatother
\def\comment#1{}

\newcommand{\stkout}[1]{
	\color{red}\ifmmode\text{\sout{\ensuremath{#1}}}\else\sout{#1}\fi\color{black}}

\IEEEoverridecommandlockouts

\begin{document}
% \receiveddate{XX Month, XXXX}
% \reviseddate{XX Month, XXXX}
% \accepteddate{XX Month, XXXX}
% \publisheddate{XX Month, XXXX}
% \currentdate{\today}
% \doiinfo{OJCOMS.2024.011100}

% \markboth{}{Lancho, Weiss {et al.}}

\title{RF Challenge: The Data-Driven Radio Frequency Signal Separation Challenge}

\author{Alejandro Lancho\authorrefmark{1,2,*,‡}, Member, IEEE, Amir Weiss \authorrefmark{3,*,‡}, Senior Member, IEEE, Gary C.F. Lee\authorrefmark{4,*}, Member, IEEE, Tejas Jayashankar\authorrefmark{5}, Student Member, IEEE, Binoy G. Kurien\authorrefmark{6}, \\ Yury Polyanskiy\authorrefmark{5}, Fellow, IEEE, Gregory W. Wornell\authorrefmark{5}, Fellow, IEEE}
% Department of Electrical Engineering and Computer Science, 
\affil{Universidad Carlos III de Madrid, Leganés 28911 Spain}
\affil{ Gregorio Marañón Health Research Institute, Madrid 28007 Spain}
\affil{Bar-Ilan University, Ramat Gan, Israel}
\affil{Institute for Infocomm Research, 138632 Singapore}
\affil{Massachusetts Institute of Technology, Cambridge, MA 02139 USA}
\affil{MIT Lincoln Laboratory, Lexington, MA 02421 USA}
\corresp{Corresponding author: Alejandro Lancho (email: alancho@ing.uc3m.es).}
\authornote{Research was supported, in part, by the United States Air Force Research Laboratory and the United States Air Force Artificial Intelligence Accelerator under Cooperative Agreement Number FA8750-19-2-1000. The views and conclusions contained in this document are those of the authors and should not be interpreted as representing the official policies, either expressed or implied, of the United States Air Force or the U.S. Government. The U.S. Government is authorized to reproduce and distribute reprints for Government purposes notwithstanding any copyright notation herein.\\
The authors acknowledge the MIT SuperCloud and Lincoln Laboratory Supercomputing Center for providing HPC resources that have contributed to the research results reported within this paper.\\
Alejandro Lancho has received funding from the Comunidad de Madrid's 2023 Cesar Nombela program under Grant Agreement No.~2023-T1/COM-29065, from the Comunidad de Madrid under Grant Agreement No. TEC-2024/COM-89, and from the Ministerio de Ciencia, Innovación y Universidades, Spain, under Grant Agreement No.~PID2023-148856OA-I00.\\
This work is also supported, in part, by the National Science Foundation (NSF) under Grant No.~CCF-2131115.\\
The material in this paper was presented in part at the IEEE Int. Workshop Mach. Learn. Signal Process. (MLSP), Aug. 2022, the IEEE Glob. Commun. Conf. (GLOBECOM), Dec. 2022, the IEEE Int. Conf. Acoust., Speech, Signal Process. (ICASSP), Jun. 2023, and the IEEE Int. Conf. Acoust., Speech, Signal Process. (ICASSP), Apr. 2024.\\
*G. Lee and A. Lancho and A. Weiss were formerly with MIT.\\
‡ Equal contribution.}

\begin{abstract}
We address the critical problem of interference rejection in radio-frequency (RF) signals using a data-driven approach that leverages deep-learning methods. A primary contribution of this paper is the introduction of the RF Challenge, which is a publicly available, diverse RF signal dataset for data-driven analyses of RF signal problems. Specifically, we adopt a simplified signal model for developing and analyzing interference rejection algorithms. For this signal model, we introduce a set of carefully chosen deep learning architectures, incorporating key domain-informed modifications alongside traditional benchmark solutions to establish baseline performance metrics for this intricate, ubiquitous problem. Through extensive simulations involving eight different signal mixture types, we demonstrate the superior performance (in some cases, by two orders of magnitude) of architectures such as UNet and WaveNet over traditional methods like matched filtering and linear minimum mean square error estimation. Our findings suggest that the data-driven approach can yield scalable solutions, in the sense that the same architectures may be similarly trained and deployed for different types of signals. Moreover, these findings further corroborate the promising potential of deep learning algorithms for enhancing communication systems, particularly via interference mitigation. This work also includes results from an open competition based on the RF Challenge, hosted at the 2024 IEEE International Conference on Acoustics, Speech, and Signal Processing (ICASSP'24).

% Through extensive simulations involving eight different signal mixture types, we demonstrate the superior performance (in some cases, of two orders of magnitude) of architectures such as UNet and WaveNet over traditional methods like matched filtering and linear minimum mean square error estimation. Our findings from the RF Challenge reveal the promising potential of deep learning algorithms for enhancing communication systems, particularly via interference mitigation.}
\end{abstract}

\begin{IEEEkeywords}
Interference rejection, deep learning, source separation, wireless communication.
% Enter key words or phrases in alphabetical order, separated by
% commas. Using the \textit{IEEE Thesaurus} can help you find the best
% standardized keywords to fit your article. Use the \underline{\href{https://www.ieee.org/publications/services/thesaurus.html}{thesaurus
% access request form}} for free access to the \textit{IEEE Thesaurus}.
\end{IEEEkeywords}

%\IEEEspecialpapernotice{(Invited Paper)}

\maketitle
\section{INTRODUCTION}

\IEEEPARstart{T}{he} rapid proliferation of wireless technologies is driving an increasingly congested radio spectrum. Emerging services, such as virtual reality and augmented reality, demand substantial bandwidth to operate effectively~\cite{qualcomm_AR}. Concurrently, new applications for \gls{urllc} and \gls{mtc} are imposing strict requirements on reliability, latency, and energy efficiency. These requirements necessitate advanced interference management strategies that go beyond traditional resource orthogonalization in time and frequency domains. As a result, wireless systems must adopt sophisticated interference management techniques to support these diverse, coexisting demands on the spectrum~\cite{Hirzallah17,Naik21}.

Standard solutions for this ubiquitous problem involve filtering out interference by masking irrelevant parts of the time-spectrum grid or using multi-antenna capabilities to focus on specific spatial directions. In this paper, however, we focus on the case where the interference overlaps both in time and frequency with the \gls{soi}, and there is no spatial diversity to be exploited. Such a challenging case can occur, for example, in single-antenna devices or multi-antenna devices with insufficient spatial resolution to satisfactorily spatially filter the interference.\footnote{While we do not include technical details, it can be shown that the multi-antenna case can be effectively equivalent to a single-channel case after applying a beamforming vector to multivariate data from an array.} In such situations, judicious and effective solutions would have to exploit the specific underlying statistical structure of the interference, potentially via learning techniques.

Throughout this paper, we will also adopt the common terminology of \emph{co-channel} interference to refer to other waveforms that operate at the same time and the same frequency band as the \gls{soi}~\cite{oyedare2022interference}. Such co-channel interference can be reduced by the use of \emph{interference mitigation} techniques, often via \emph{signal separation} methods.\footnote{We will henceforth refer to signal separation also as source separation or \emph{interference rejection}, interchangeably. Furthermore, the interference rejection problem can also be understood as a denoising problem, where we aim to remove the \gls{soi} from the ``noise", which, in this case, is the non-Gaussian interference signal.} In this context, the goal is to extract the \gls{soi} with the highest possible fidelity, thereby enhancing downstream task performance (e.g., detection, demodulation, and decoding).
\subsection{PREVIOUS WORK}\label{sec:prevwork}
The simplest solution for interference rejection in communication systems is
to filter the received signal using a \emph{matched filter} that is matched to the one used to
generate the baseband signal waveform at the transmitter~\cite{vantrees_01}, thereby implicitly
(and most likely incorrectly) treating the interference as \gls{awgn}. Perhaps surprisingly, this
is often the \emph{only} interference mitigation method employed in existing wireless
communication systems. However, it is well-known that the matched filter solution, while
guaranteed to be optimal (in the maximum \gls{snr} sense) for an \gls{awgn} channel~\cite{vantrees_01}, is
certainly not necessarily optimal in other settings. Consider, for example, an interference that is a
communication signal generated from another communication system, overlapping with the \gls{soi}
in time and frequency. In this case, in addition to (Gaussian) noise, the received signal will be contaminated with a non-Gaussian interference as well. In this scenario, matched filtering is likely to be suboptimal, thus creating the possibility for other source separation techniques to provide performance gains.

There are, indeed, various source separation methods in the literature that were proposed and specifically designed for digital communication signals. One noteworthy approach is maximum likelihood sequence estimation of the target signal, for which algorithms such as particle filtering \cite{tu2007particle} and per-survivor processing algorithms \cite{tu2008single} can be used. However, methods such as maximum likelihood, often referred to as ``model-based" methods, require prior knowledge of the statistical models of the relevant signals, which may not be known or available in practical scenarios. As a result, in practice these methods are often suboptimal, and in some cases perform poorly (see, e.g., \cite{lee2011interference,chevalier2018third}).

In those cases where the statistical models of the involved signals are not fully known, a more realistic (though challenging) paradigm is to assume that only a dataset of the underlying communication signals is available. This can be obtained, for example, through direct/background recordings or using high-fidelity simulators (e.g., \cite{o2016radio}), allowing for a \emph{data-driven} approach to source separation. In this setup, \glspl{dnn} arise as a natural choice. This data-driven version of the source separation problem has been promoted within the context of the ``RF Challenge'' dataset, where signal separation with little to no prior information is pursued~\cite{rfchallenge}.

While \gls{ml} techniques have shown promise in source separation within the vision and audio domains~\cite{stoller2018wave,nugraha2016multichannel}, the \gls{rf} domain presents unique challenges. Typically, these methods  exploit domain-specific knowledge relating to the signals' characteristic structures. For example, color features and local dependencies are useful for separating natural images~\cite{gandelsman2019double}, whereas time-frequency spectrogram masking methods are commonly adopted for separating audio signals~\cite{huang2015joint}. In contrast to natural signals, such as images or audio recordings, most \gls{rf} signals are different in nature: i) they are \emph{synthetically} generated via digital signal processing circuits; ii) they originate from discrete random variables;\footnote{This is important because it implies that communication signals follow non-differentiable probability mass functions rather than differentiable probability density functions, which makes learning such distributions more challenging. These numerical challenges are further discussed in \cite{jayashankar2023score-neurips}.} and iii) they typically present an intricate combination of short and long temporal dependencies. On top of these differences, the mixture signals may overlap in time and frequency. All this together implies that classical, ``handcrafted" model-based solutions---while successful in other domains---may fail in the \gls{rf} signal domain (e.g., \cite{lee2023neural,github1}). %Moreover, \gls{ml}-based solutions often require large datasets, and while these are abundantly available and easily accessible in the vision and audio domains, they are still scarce in the \gls{rf} domain.
\subsection{THE NEED FOR RF SIGNAL DATASETS}
\gls{ml}-based solutions often require large datasets. While large datasets are readily available in domains like vision and audio, they remain scarce in the \gls{rf} domain, despite the widespread importance of digital \gls{rf} communication signals in our everyday lives (\hspace{-0.005cm}\cite[Ch.\ 2.4]{oyedare2024comprehensive} and references therein).

Among the available datasets, notable examples include the one provided by DeepSig, which offers several synthetically generated signals from GNU Radio for modulation detection and recognition~\cite{rfdeepai}, and those available through IQEngine, a web-based \gls{sdr} toolkit for analyzing, processing, and sharing \gls{rf} recordings~\cite{iqengine}.

A relatively new dataset designed specifically for source separation of RF signals is the ``RF Challenge"~\cite{rfchallenge}. This dataset includes several raw \gls{rf} signals with minimal to no information about their generation processes. 
% Within the RF Challenge, the \emph{single-channel signal separation challenge} focuses on two goals:
% \begin{enumerate}
%     \item Separate a \gls{soi} from the interference;
%     \item Demodulate the (digital) \gls{soi} component in such a mixture.
% \end{enumerate}
The lack of prior knowledge of the signal structure, combined with the possible complete overlap in time and frequency between the constituent signals, renders conventional separation via classical, and in particular linear, filtering techniques ineffective. Addressing this challenge calls for new learning methods and architectures \cite{lee2022exploiting,jayashankar2023score-neurips} that must go beyond the state of the art. The RF Challenge was created to promote the development of solutions to important problems particular to the \gls{rf} domain, similar to how datasets such as MNIST, ImageNet, VAST, and HPC Challenge (\hspace{-0.005cm}\cite{lecun1998gradient,deng2009imagenet,cook2014vast,luszczek2005introduction}, respectively) have catalyzed research considerably in their respective areas by creating \emph{standard benchmarks} and high-quality data.%In particular, they need to implicitly identify less obvious features, which are specific and unique to RF digital communication signals, that are not readily discernible through conservative time and/or frequency domain analysis.

%This paper focuses on the signal datasets provided by the RF Challenge. 
The data associated with the RF Challenge are publicly available at \url{https://rfchallenge.mit.edu/datasets/} and contain several datasets of \gls{rf} signals recorded over the air or generated in lab environments. Specifically, most signals in the dataset are from the 2.4 GHz industrial, scientific, and medical (ISM) band. Only the 5G signals in the dataset were generated using a cable setup and a simulated 5G channel environment. %Specifically, the data were recorded in the \gls{ism} radio bands with the ultimate goal of improving coexistence among WiFi, Bluetooth, ZigBee, and other \gls{ism} band users.
\subsection{CONTRIBUTIONS}\label{sec:contributions}
Our main contribution in this paper is the comprehensive presentation of the RF Challenge dataset for the \emph{single-channel signal separation challenge}, focusing on two goals:
\begin{enumerate}
    \item Separate a \gls{soi} from the interference;
    \item Demodulate the (digital) \gls{soi} in such a mixture.
\end{enumerate}
Rather than considering the classical formulation of source separation, we tackle this problem from a fresh, data-driven perspective. Specifically, we introduce a novel \gls{ml}-aided approach to signal processing in communication systems, leveraging data-driven solutions empowered by recent advancements in deep learning techniques. These solutions are made feasible by progress in computational resources and the publicly available signal datasets we created and organized. We highlight that the methods developed within this research domain not only enable \gls{rf}-aware \gls{ml} devices and technology, but also hold the potential to enhance bandwidth utilization efficiency, facilitate spectrum sharing, improve performance in high-interference environments, and boost system robustness against adversarial attacks.

Through an extensive presentation of results, % stemming from our efforts,  in recent years, 
we show the potential of data-driven, deep learning-based solutions to significantly enhance interference rejection, and achieve improvements by orders of magnitude in both \gls{mse} and \gls{ber} compared to traditional signal processing methods. To support this claim, we introduce two deep learning architectures that we have established as benchmarks for interference mitigation, along with the performance results of the top teams from the ``Data-Driven Radio Frequency Signal Separation Challenge" that we  hosted at the ICASSP'24 Signal Processing Grand Challenges~\cite{datadrivenrf2024}. 

Finally, we conclude the paper by outlining a series of future directions focused on mitigating non-Gaussian interference in wireless communication systems. We expect these research directions, reinforced by competitions such as the recent SP Grand Challenge at ICASSP'24, to gain increasing relevance in the near future, and we invite researchers worldwide to actively contribute to advancing this field.
\vspace{-0.3cm}
\subsection{Notations}
We use lowercase letters with standard font and sans-serif font, e.g., $x$ and $\rndx$, to denote deterministic and random scalars, respectively. Similarly, $\vecx$ and $\rvecx$ represent deterministic and random vectors, and $\matX$ and $\rmatX$ denote deterministic and random matrices. Additionally, $\rndx[n]$ is used to represent the $n$-th random sample of the vector-form random signal $\rvecx$. 
The uniform distribution over a set $\setS$ is denoted as ${\rm Unif}(\setS)$, and for $K\in\naturals$, we define $\setS_{1:K}\triangleq\{1,\ldots,K\}$. For brevity, we refer to the complex normal distribution as Gaussian. We denote $\matC_{\rndz\rndw}\triangleq\Exop\left[\rvecz\herm\rvecw\right]\in\complexset^{N_z \times N_w}$ as the cross-covariance matrix of the zero-mean vectors $\rvecz\in\complexset^{N_z\times 1}$ and $\rvecw\in\complexset^{N_w\times 1}$ (specializing to the auto-covariance $\matC_{\rndz\rndz}$ when $\rvecz=\rvecw$). The indicator function $ \mathbbm{1}_{\setE}$ returns $1$ when the event $\setE$ occurs, and $0$ otherwise.
%%%%%
\vspace{-0.25cm}
\section{PROBLEM STATEMENT}\label{sec:prob_statement}
\begin{figure*}[t!]
\centering
\begin{subfigure}[t]{\columnwidth}
\centering
\includegraphics[width=\columnwidth]{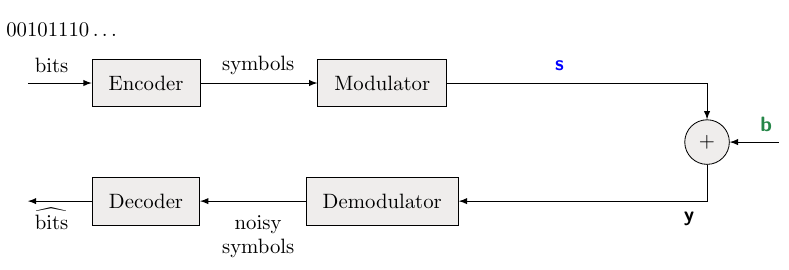}
\caption{Traditional communication scheme with no specific block for interference mitigation.}
\label{fig:comm_sch_no_mit}
\end{subfigure}
\begin{subfigure}[t]{\columnwidth}
\centering
\includegraphics[width=\columnwidth]{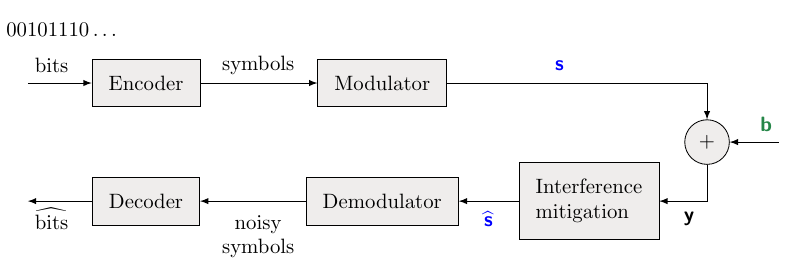}
\caption{Communication scheme with a dedicated building block for interference mitigation.}
\label{fig:comm_sch_mit}
\end{subfigure}
\caption{Communication schemes considered in this work.}
\label{fig:comm_sch}
\end{figure*}
We consider the point-to-point, single-channel,\footnote{The single-channel model encompasses scenarios such as single-antenna links and multi-antenna links where the spatial resolution is insufficient, resulting in a single effective channel between the transmitter and receiver.} baseband signal model depicted in Fig.~\ref{fig:comm_sch}, where a transmitter aims to communicate a signal that carries a stream of encoded and modulated bits, referred to as the \gls{soi} and denoted as $\rvecs$. The signal is measured at the intended receiver in the presence of an unknown interference signal, denoted as $\rvecb$. The ultimate goal of the receiver is to successfully detect (or recover) the transmitted bits (or message) with the highest possible reliability, measured by the \gls{ber}.

The input-output relation for a received, sampled, discrete-time baseband signal of length $N$ samples is given by
\begin{equation}\label{eq:mixturemodel}
    \rvecy = \rvecs + \rvecb \;\in \mathbb{C}^{N\times1}.
\end{equation}

This simplified model allows us to focus solely on the problem of interference rejection and the potential contributions of \gls{ml} in this context. One can consider this model as the resulting input-output relation after successfully completing crucial processing stages in a communication system, such as time synchronization, channel estimation, and equalization. Although these aspects are deferred for future research, we acknowledge their importance in ensuring the correct operation of any practical communication system. Nonetheless, as we shall demonstrate throughout this paper, studying this building block in isolation enables us to understand the potential impact and challenges of integrating \gls{ai} capabilities into \gls{rf} communication receivers.

Furthermore, we consider the case where the generation process of the interference signal $\rvecb$ is unknown. Specifically, we assume that the interference consists of an unknown \gls{rf} signal originating from another system operating in the same time-frequency band, possibly contaminated by \gls{awgn}.

Recall that we focus on digital communication signals as \gls{soi} in this work. In digital communication systems, the ultimate goal is to reliably recover the transmitted bits (or messages). Therefore, we consider the \gls{ber} as a central measure of performance in this paper.

Note that we consider a scenario with non-Gaussian interference of unknown generation process, for which the optimal solution to minimize the \gls{ber} is generally unknown.\footnote{If the interference were Gaussian, applying a matched filter at the receiver, which is matched to the one used to modulate the encoded bits prior to decoding, would be optimal for the \gls{ber} criterion (see Sec.~\ref{sec:trad_methods}).} Under this setting, various receiver architectural designs can be devised based on different principles, aiming to achieve the best possible \gls{ber} performance. In this work, we propose a hybrid, "smart" receiver that first performs interference mitigation in a data-driven manner using a \gls{dnn}. This approach aims to learn the relevant features of the unknown interference signal, as well as their statistical interactions with the relevant features of the \gls{soi}, in order to mitigate it. Then, by treating the residual interference as Gaussian noise, we apply standard matched filtering prior to decoding, so as to increase the postprocessing \gls{snr}. Consequently, we introduce a second measure of performance, namely the \gls{mse} between the estimated \gls{soi} $\hat{\rvecs}$ and the true transmitted \gls{soi} $\rvecs$, to assess the signal quality after interference rejection and before decoding.\footnote{Other performance measures could be considered depending on the specific receiver designs under consideration. Examples include packet-error rate when channel coding is part of the pipeline, peak-SNR to evaluate the fidelity of the reconstructed signal’s amplitude, and outage probability, which is especially relevant for systems requiring guaranteed quality of service, such as \gls{urllc}.}
\subsection{SIGNAL MODELS}
In this section, we categorize the various types of signals considered in this work based on our knowledge of their generation process.

When the signal's generation process is known, we have detailed information about the generated
signal. Specifically, we can generate a synthetic dataset of signals for the sake of learning a
data-driven source separation module. This approach is valuable when model-based solutions are
infeasible, either because the model of the interference is unknown (but the \gls{soi}'s model is
known) or because the signal models are analytically intractable or such that lead to computationally impractical solutions.
We will further categorize signals with a known generation process into single-carrier and
multi-carrier signals.

When the signal generation process is unknown, we assume that we have datasets available, obtained through recordings or high-fidelity simulations. Thus, any knowledge relevant to performing source separation on these types of signals must be learned from the data.
\subsubsection{SIGNALS WITH A KNOWN GENERATION PROCESS}
We consider single-carrier and multi-carrier signals generated by linearly modulating symbols from constellations in the complex plane. Signals generated in this manner correspond to a prevalent class of digital communication signals observed in typical \gls{rf} frequency bands.

\begin{figure}[t!]
    \centering
    \raisebox{12mm}{\includegraphics[width=0.8\columnwidth]{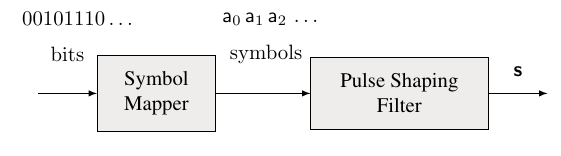}}\vspace{-0.75cm}
    \caption{Block diagram for the generation process of the single-carrier signal, which modulates bits that are mapped into symbols from a complex-valued constellation before being filtered using a given pulse shaping filter.}
    \label{fig:sc-diag}
\end{figure}

\paragraph*{Single-Carrier Signals}
We consider single-carrier signals bearing $M$-bit long messages, which are mapped to $L$ symbols from a given complex constellation (e.g., \gls{qpsk}) using Gray coding. The bits are randomly generated via a fair coin toss and are all \gls{iid}. The $n$-th sample of $\rvecs\in\mathbb{C}^{N\times1}$ is expressed as
\begin{equation}\label{eq:singlecarrier}
\rnds[n] = \sum_{\ell=0}^{L-1} \rnda_\ell \cdot g[n - \ell F - \tau_0],
\end{equation}
where $\rnda_\ell\in\setA$ denotes a complex discrete symbol to be transmitted, with $\setA$ being the constellation of (possibly complex-valued) symbols, $F\in\integers$ is the symbol interval (in discrete-time), $\tau_0\in\setS_{0:F-1}$ is the offset for the first symbol, and $g[n]$ is the discrete-time impulse response of the transmitter filter (pulse shaping function). Figure~\ref{fig:sc-diag} shows a simplified diagram for the generation process of the considered single-carrier signal type.
\vspace{-0.3cm}
\paragraph*{Multi-Carrier Signals}
For multi-carrier signals, we focus on \gls{ofdm} signals, which are among the most widely used in key wireless communication technologies such as 5G and WiFi. 
An \gls{ofdm} signal consists of $K$ orthogonal subcarriers, each carrying a symbol from a given (generally complex-valued) constellation \cite{hwang2008ofdm}. We will consider the \gls{qpsk} constellation for the numerical examples of this paper.

\begin{figure}[t!]
    \centering
    \includegraphics[width=1\linewidth]{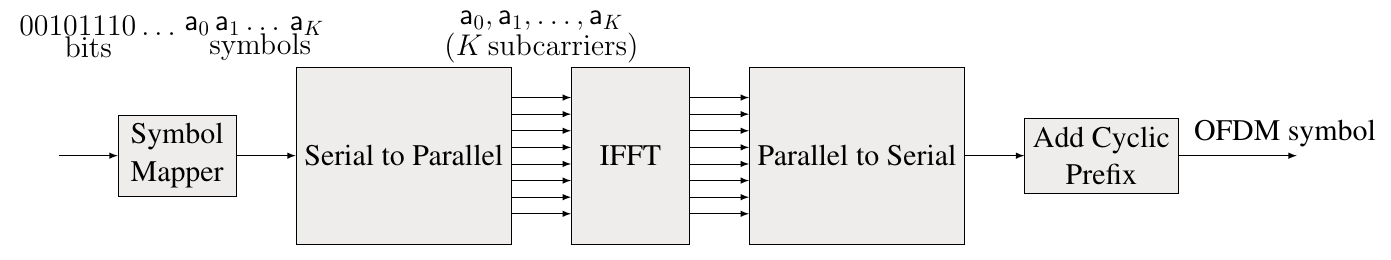}
    \caption{Block diagram for the generation process of an OFDM symbol carrying symbols in each active subcarrier.}
    \label{fig:OFDMsymbdiag}
\end{figure}

In this case as well, the bits are randomly generated using a fair coin toss in an \gls{iid} manner and then mapped to symbols from the given constellation using Gray coding. The $n$-th sample of the \gls{soi} $\rvecs\in\mathbb{C}^{N\times1}$ is given by
\begin{multline}\label{eq:ofdmform}
    \rnds[n] = \sum_{p=0}^{P-1}\sum_{k=0}^{K-1} \rnda_{k,p} \, r[n-p\cdot(K+\Tcp)-\Tcp, \, k],
\end{multline}
where
\begin{equation}\label{eq:indicatorandexp}
    r[n, \, k] \triangleq \exp\left({j 2\pi kn/K}\right)  \cdot \mathbbm{1}_{\left\{-\Tcp \leq n < K\right\}}.
\end{equation}
Here, $K$ represents the total number of orthogonal complex exponential terms (subcarriers), where not all of them are necessarily active.\footnote{An ``active" subcarrier is one that is being used to convey information, not necessarily random (e.g., pilots for the sake of channel estimation, and recall that pilots are predetermined, deterministic and known).} The value of $K$ corresponds to the \gls{fft} size of the inverse \gls{fft} (IFFT) involved in generating an \gls{ofdm} signal. The coefficients $\rnda_{k,p}\in\mathcal{A}$ are the information modulating symbols, where $\mathcal{A}$ represents the constellation.\footnote{For simplicity, the constellation includes the zero symbol, so \eqref{eq:ofdmform} accounts for inactive subcarriers as well.} A \gls{cp} is typically added before an \gls{ofdm} symbol. Thus, each \gls{ofdm} symbol is described within the interval $[-\Tcp, K]$, where $\Tcp$ is the \gls{cp} length. The signals then span $P=N/(K+\Tcp)\in\integers$ \gls{ofdm} symbols, and their individual finite support is reflected by the finitely-supported function $r[n,k]$ in \eqref{eq:indicatorandexp}. Figure~\ref{fig:OFDMsymbdiag} illustrates the block diagram of the \gls{ofdm} symbol generation process.
\vspace{-0.3cm}

\subsubsection{SIGNALS WITH UNKNOWN GENERATION PROCESS}
When dealing with an unknown interference (e.g., from a different technology), accessing the signal generation process, which could potentially allow for the design of specific interference rejector, is often rare. However, one can rely on recorded interference signals to \emph{learn} how to design the interference rejector from the data. Another scenario where we may lack access to the generation process but still need to separate signals is the classical blind source separation problem. In this case, we may not know any of the signal models involved in the communication process, and our goal is simply to separate the superimposed signals into their constituent components. Finally, we could also consider the case where the generation process of the signals is known but is too complicated for deriving analytical solutions.

In any of these cases, the availability of signal datasets enables the design of data-driven source separators. We note that within the RF Challenge, there is a dataset of interference signals whose generative models are unknown, which can be used to develop learned, data-driven solutions.

% The collection of datasets of relevant \gls{rf} communication signals is therefore an essential component in the development of data-driven algorithmic solutions in the context of wireless communications. Although this work focuses on interference rejection problems, these signal datasets could be utilized for other purposes, importantly, for advancing solutions to various other issues associated with communication systems. We will further discuss this in Section~\ref{sec:outlook}.
%
\section{METHODS}\label{sec:methods}
In this section, we review the methods used to perform interference rejection for the various combinations of \glspl{soi} and interference signals considered in this work. It is important to note that the signal models in this study are not necessarily known, making it impossible to derive theoretical performance bounds. Therefore, including a diverse set of numerically evaluated methods is essential. Alongside the proposed data-driven approaches, we include traditional methods that are widely used in both the literature and practical communication systems, providing well-established benchmarks for comparison.

\subsection{TRADITIONAL METHODS}\label{sec:trad_methods}
We now present two prevalent methods whose appeal comes from the balance between their theoretical justification—and, in fact, \emph{optimality} for common criteria under certain conditions—and their simplicity, an important virtue in practical systems.

\subsubsection{LINEAR MMSE ESTIMATION}\label{sec:lmmseestimation}
A computationally attractive approach that exploits the joint second-order statistics of the mixture \eqref{eq:mixturemodel} and the \gls{soi} is optimal \gls{mmse} \emph{linear} estimation. Assuming $\det(\matC_{\rndy\rndy})\neq0$ and that $\rvecs$ and $\rvecb$ are uncorrelated, the \gls{lmmse} estimator \cite{vantrees_01}, given by
\begin{equation}\label{lmmsedef}
    \widehat{\rvecs}_{\lmmse}\triangleq \matC_{\rnds\rndy}\inv{\matC_{\rndy\rndy}}\rvecy=\matC_{\rnds\rnds}\inv{\left(\matC_{\rnds\rnds}+\matC_{\rndb\rndb}\right)}\rvecy\in\complexset^{N\times 1},
\end{equation}
is constructed using the second-order statistics of the mixture that inherently take into account the potentially nontrivial temporal structure of the interference expressed through $\matC_{\rndb\rndb}$. In other words, if $\matC_{\rndb\rndb}$ somehow deviates from a scaled identity matrix, temporal cross-correlations exist.

While \eqref{lmmsedef} coincides with the \gls{mmse} estimator when $\rvecy$ and $\rvecb$ are jointly Gaussian, it is generally suboptimal due to the linearity constraint. In our case, the signal $\rnds[n]$ is a digital communication signal and is certainly not Gaussian. As for $\rndb[n]$, its statistical model is assumed to be unknown throughout the design process of the interference mitigation module. Still, it would also typically be non-Gaussian, even if it contains \gls{awgn}, which is highly plausible.

Despite \eqref{lmmsedef} not being the \gls{mmse} estimator in our scenarios of interest, it is still an important benchmark since it constitutes an attractive method for two main reasons. First, it is linear, and therefore fast and easy to implement for moderate values of $N$. Second, it \emph{only} requires knowledge of second-order statistics, which are relatively easy to accurately estimate from data, even in real-time systems. We therefore use it as one of our benchmarks whenever it is computationally feasible.\footnote{For nonstationary input signals, the required inversion of $\matC_{\rndy\rndy}$ is computationally impractical at high dimensions—matrix inversion (without a particular structure to be exploited) is generally of complexity $\mathcal{O}(N^{3})$.}

\subsubsection{MATCHED FILTERING}\label{sec:mathchedfiltering}
Matched filtering, perhaps one of the most commonly used techniques in the signal processing chain of communication systems, exploits prior knowledge about the signal waveform (only) for enhanced detection of the transmitted symbols. When the residual (additive) component is Gaussian, it is optimal in the sense that it maximizes the SNR, and it is therefore also optimal in terms of minimum \gls{ber}.

If the transmitted signal is represented by $\rnds[n] = \rnda_0 \cdot g[n]$, where $g[n]$ is the pulse shaping filter, then the matched filter would be $h_{\mf}[n]=g^*[-n]$. In practical scenarios, the pulse $g[n]$ has finite duration. After performing the complex conjugation and time-reversal to obtain $g^*[-n]$, the resulting signal is shifted appropriately to ensure causality. This shift corresponds to aligning the start of the pulse with the beginning of the observation window.

This method is also an important benchmark as it is probably still the most commonly used method for symbol detection, which is the natural choice when the residual component, be it noise or interference, is treated as \gls{awgn}.

To conclude this section, we note in passing that the (theoretically na\"ive) option of not applying an interference mitigation method, namely only applying a matched filter to the received signal~\eqref{eq:mixturemodel}, will also be considered in our simulation as a benchmark. Indeed, with the complete absence of prior knowledge of the statistical model of the interference, this plain option of simply ignoring the interference may, after all, be chosen for practical considerations. While we do not advocate for such a solution approach, we acknowledge it as a realistic (even if not a leading) benchmark.

\subsection{DATA-DRIVEN METHODS}
\label{sec:data-driven}
\begin{figure*}[t!]
\centering
\includegraphics[width=\textwidth]{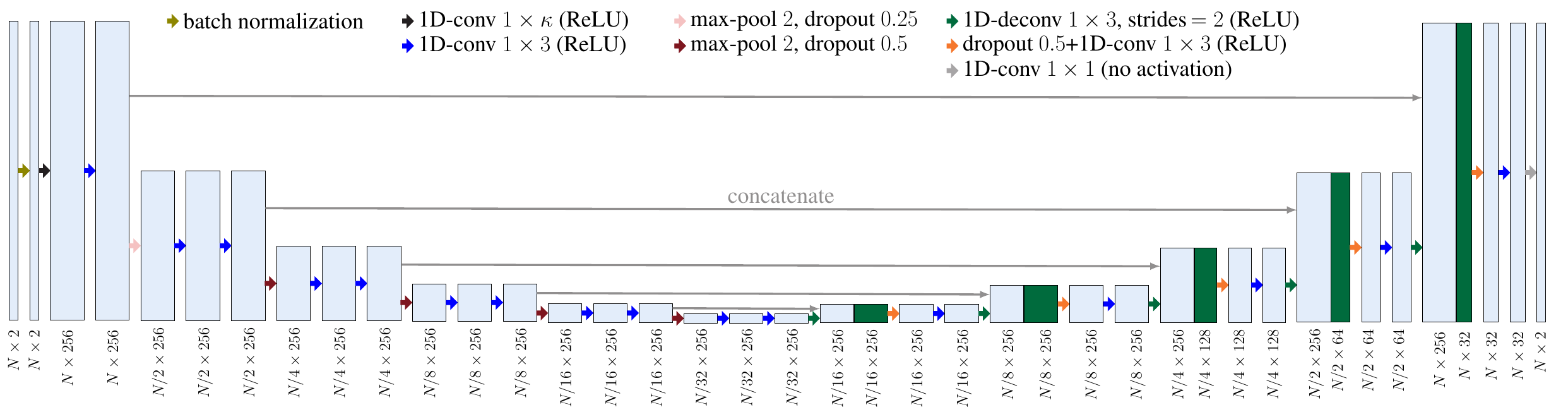}
\caption{The UNet \gls{dnn} architecture proposed for single-channel source separation of communication signals.
The parameter $\kappa$ denotes the kernel size of the first layer.}
\label{fig:unet}
\end{figure*}
This section presents the two most effective architectures we identified for data-driven source separation of \gls{rf} signals: UNet and WaveNet. This selection was informed by insights from prior work. Specifically, in \cite{lee2022exploiting} we analyzed cyclostationary Gaussian signals to in order to isolate the effect of temporal correlations (i.e., second-order statistics) from higher-order statistics, and determine key modifications to standard deep learning architectures. This enabled us to compute the (exact) optimal \gls{mmse} solution, which was computable in this setup. We observed that extending the kernel size in the initial layer of the UNet—proportional to the effective length of the cross-correlation between the signals—significantly improved performance, closely approximating the optimal \gls{mmse} estimator in the cyclostationary Gaussian signal scenario. These findings guided our modifications of the UNet architecture, enhancing its performance when applied to real-world \gls{rf} signals from the RF Challenge dataset.

Additionally, in \cite{lee2023neural} we evaluated state-of-the-art architectures from the audio domain on scenarios involving superimposed \gls{ofdm} signals, now addressing separation based on higher-order statistics, where theoretically perfect separation was possible. However, these architectures struggled without domain-specific modifications. To address this, we introduced structural changes that led to up to a 30 dB improvement in separation performance. These modifications involved extending the kernel size in UNet-like architectures, so as to align with the OFDM's underlying FFT size, and employing dilated convolutions, which also motivated the inclusion of WaveNet in our subsequent works \cite{jayashankar2023score-neurips,datadrivenrf2024}.

Throughout the course of our research, we evaluated additional architectures, many of which did not yield notable improvements. Comparative performance results for these models are available in a dedicated GitHub repository for further reference.\footnote{\url{https://github.com/RFChallenge/SCSS_DNN_Comparison}}

We henceforth assume we have a dataset of $D$ \gls{iid} copies of $\{(\rvecy^{(i)},\rvecs^{(i)})\}_{i=1}^D$, i.e., the baseband versions of the mixture and SOI, whose real and imaginary parts are their in-phase and quadrature components, respectively.

\subsubsection{UNET}\label{sec:unet}
The UNet, as depicted in Fig.~\ref{fig:unet}, is a type of \gls{dnn} originally proposed for biomedical image segmentation \cite{ronneberger2015unet}. Its versatility has led to its adoption in various other applications, including spectrogram-based \gls{rf} interference cancellation \cite{akeret2017radio} and audio source separation \cite{stoller2018wave, tzinis2020sudo}. These applications typically correspond to a multivariate regression setup with identical dimensions for both input and output data.

Similarly to these aforementioned works, our approach employs 1D-convolutional layers to better capture the temporal features of time-series data. To effectively handle (baseband) complex-valued signals, inspired by widely linear estimation techniques \cite{picinbono1995widely}, we represent the real and imaginary parts as separate input channels. The UNet architecture comprises downsampling blocks, which operate on progressively coarser timescales, and incorporates skip connections to combine features from different timescales with the upsampling blocks.

It is well known that the careful design of a neural network architecture, tailored to the specific application, can significantly impact performance, as demonstrated by our experiments and architectural choices. 
Specifically, unlike standard CNN-based architectures tailored for image processing, which originally employed short kernels of size $3$ in all layers, our UNet architecture features a first convolutional layer with a nonstandard, comparatively long kernel (indicated by $\kappa$ in Fig.~\ref{fig:unet}), which is of size $101$—a difference of two orders of magnitude. We observed that proper adjustment of this hyperparameter to capture the effective correlation length of both the \gls{soi} and interference facilitates (and perhaps enables, as some of our findings indicates) the extraction of additional long-scale temporal structures of both signals, leading to performance gains of an order of magnitude compared to the originally proposed UNet~\cite{lee2022exploiting}.\footnote{Our proposed UNet architecture for source separation of \gls{rf} signals can be found at \url{https://github.com/RFChallenge/icassp2024rfchallenge/blob/0.2.0/src/unet_model.py}.}

\subsubsection{WaveNet}
\begin{figure}[t!]
\centering
\includegraphics[width=0.95\columnwidth]{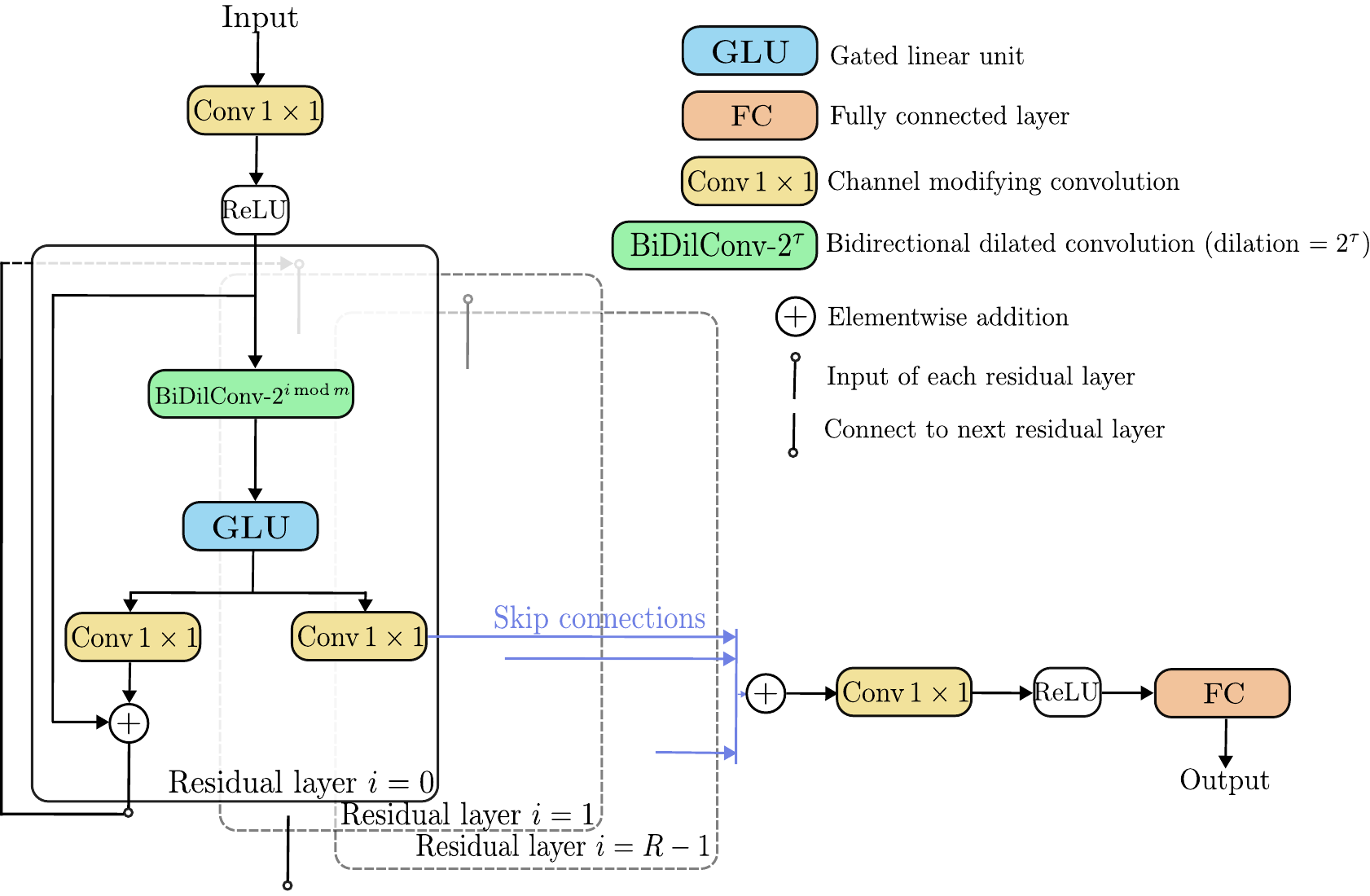}
\caption{The WaveNet \gls{dnn} architecture proposed for single-channel source separation of communication signals.}
\label{fig:wavenet}
\end{figure}
\begin{figure}[t!]
\centering
\includegraphics[width=0.95\columnwidth]{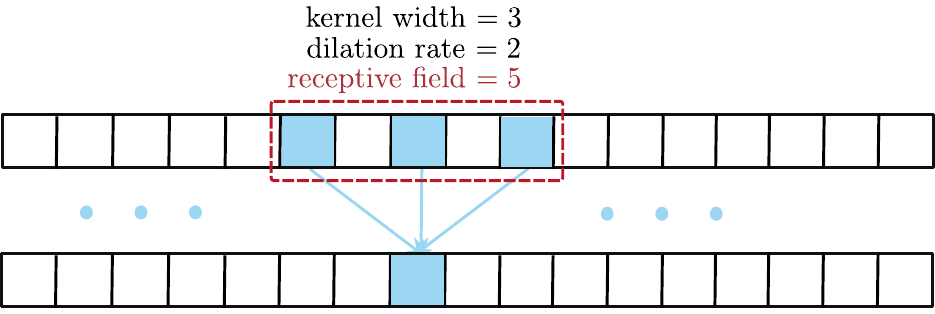}
\caption{A dilated convolution operation with a kernel width of $3$ and a dilation rate of $2$, which results in a receptive field of $5$.}
\label{fig:dilated_conv}
\end{figure}
\begin{figure*}[t]
\centering
\includegraphics[width=\textwidth]{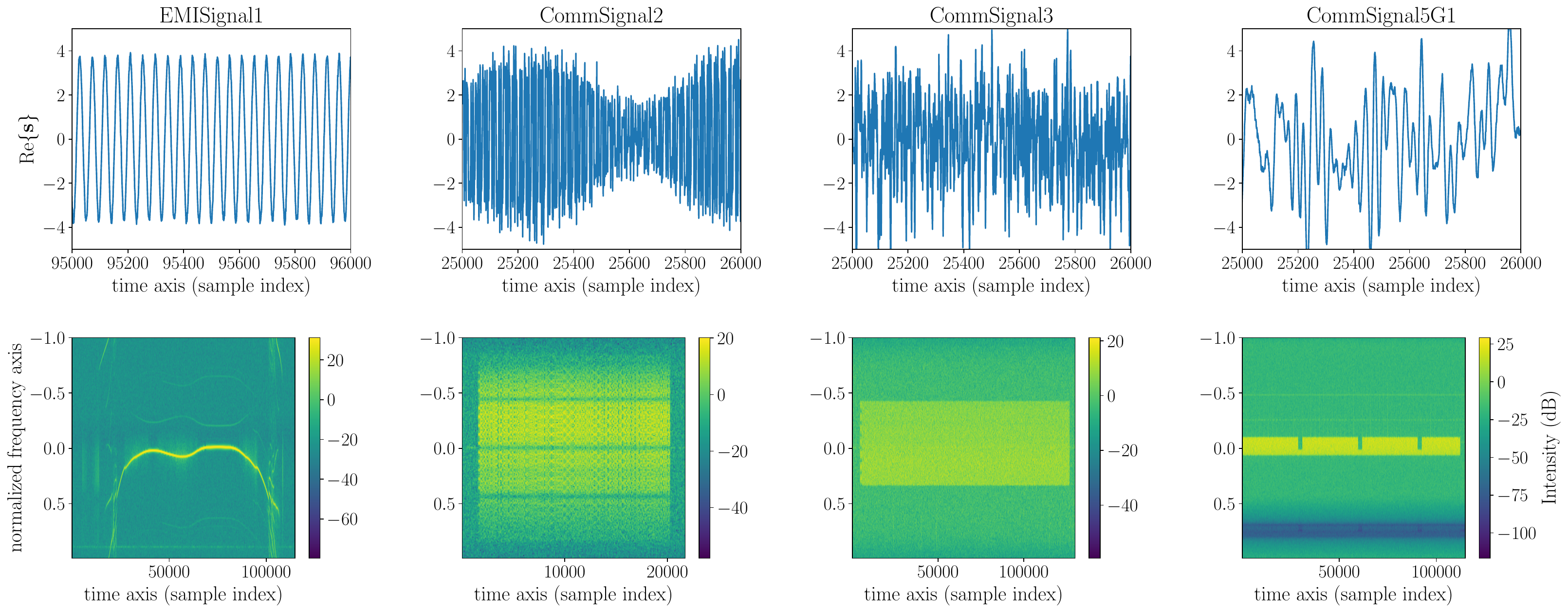}
\caption{Representative frames of the four interference signal types in the dataset: EMISignal1, CommSignal2, CommSignal3, and CommSignal5G1. Top: Real part of the waveforms, $\mathrm{Re}\{\vecs\}$; Bottom: Spectrogram of the respective signal frames.}
\label{fig:dataset}
\end{figure*}
The WaveNet architecture \cite{Oord16} was initially introduced as a generative neural network for synthesizing raw audio waveforms. In subsequent work, it was adapted for the task of speech denoising \cite{Rethage18}. At its core, the architecture uses stacked layers of convolutions with gated activation units. Unlike the downsampling and upsampling networks used in UNets,
WaveNet preserves the temporal resolution at each layer while expanding the temporal receptive field by using dilated convolutions. As shown in Fig.~\ref{fig:dilated_conv}, a dilated convolution can be interpreted as a kernel with spacing between elements, allowing the model to capture longer temporal dependencies without downsampling the sequence.
For example, a dilated convolution with a kernel width of 3 and a dilation of 2 has an effective receptive field of 5.

As illustrated in Fig.~\ref{fig:wavenet}, the WaveNet employs $R$ residual blocks with dilated convolutions, where the output of block $i-1$ serves as the input to block $i$, for $i \in \{0, \dots, R-1\}$. The dilated convolutions assist in learning long-range temporal and periodic structures. The dilations start small and successively increase, such that the dilation at block $i$ is given by $2^{i\, \textrm{mod}\, m}$, where $m$ is the dilation cycle length. For example, if the dilation periodicity is $m = 10$, then in block $i=9$ the dilation is 512, and in block $10$ the dilation is reset to 1. This allows the network to efficiently trade off between learning local and global temporal structures. All residual blocks use the same number of channels, $C$. Our WaveNet specifically uses $R=30$ residual blocks, with a dilation cycle $m=10$, and a number of channels per residual block of $C=128$.

A few key modifications were made to facilitate training with \gls{rf} signals compared to the original WaveNet \cite{Oord16}. First, since we are dealing with complex-valued continuous waveforms, we train on two-channel signals where the real and imaginary components of the \gls{rf} signals are concatenated in the channel dimension. Second, we train with an \gls{mse} (squared $\ell_2$) loss, as we did with the UNet. We monitor the validation MSE loss, and once the loss stops decreasing substantially, we stop training early. Lastly, we increased the channel dimension up to $C=128$ to learn complex \gls{rf} signals such as OFDM signals. Additionally, during data loading, we perform random time shifts and phase rotations on the interference to gain diversity and simulate typical transmission impairments in \gls{rf} systems.\footnote{Our proposed WaveNet architecture for source separation of \gls{rf} signals can be found at \url{https://github.com/RFChallenge/icassp2024rfchallenge/blob/0.2.0/src/torchwavenet.py}.} 
\section{RESULTS}
\label{sec:results_SOI_knwon}
We present a diverse set of results for \gls{rf} signal separation, examining various mixtures of signal types. Each combination exhibits unique joint statistical properties, introducing different levels of complexity in the task of learning effective signal separators.

 We compare the performance of several approaches, including data-driven, neural network-based separators, as well as more traditional, commonly used methods. Beyond showcasing our contributions in developing \gls{ml}-enhanced \gls{rf} signal separation architectures, these results are also crucial for establishing standardized benchmarks that will serve as baselines for future research in this emerging field.

To analyze decoding capabilities (in terms of \gls{ber}) alongside interference rejection capabilities (in terms of the \gls{mse} of the ``denoised'' \gls{soi}), we consider \glspl{soi} with known generative processes in this work. Specifically, we consider two different \glspl{soi} and four types of interferences, resulting in eight different combinations of mixture types, each of length $N=40,960$ samples.

For the \glspl{soi}, we have:
\begin{enumerate}
    \item \textbf{QPSK:} A single-carrier \gls{qpsk} signal with an oversampling factor of $F=16$, modulated by a root-raised cosine pulse shaping function with a roll-off factor of $0.5$ that spans $128$ samples ($8$ \gls{qpsk} symbols due to the employed oversampling factor). We further apply an offset for the first symbol of $\tau_0=8$ samples. See %Fig.~\ref{fig:psf} for a plot of the root-raised cosine pulse shaping function and 
    Fig.~\ref{fig:sc-diag} for a simplified diagram of the generation process of this \gls{soi}, which we refer to as ``QPSK''.
    \item \textbf{OFDM-QPSK:} An \gls{ofdm} signal where each subcarrier bears a \gls{qpsk} symbol. We refer to this signal as ``OFDM-QPSK''. We set $\Tcp=16$, $K=64$ subcarriers, with $56$ active subcarriers (i.e., the $8$ inactive subcarriers ``carry'' the zero symbol). Recall that these quantities were defined in \eqref{eq:ofdmform}. A simplified diagram of the generation process of this \gls{soi} is shown in Fig.~\ref{fig:OFDMsymbdiag}.
\end{enumerate}

The following four types of interference signals are only available through provided recordings, hence their generation process is unknown:
\begin{enumerate}
    \item \textbf{EMISignal1:} Electromagnetic interference from unintentional radiation from an unknown RF-emitting device with a recording bandwidth of $25$ MHz.
    \item \textbf{CommSignal2:} A digital communication signal from a commercially available wireless device with a recording bandwidth of $25$ MHz.
    \item \textbf{CommSignal3:} Another digital communication signal from a commercially available wireless device with a recording bandwidth of $25$ MHz.
    \item \textbf{CommSignal5G1:} A 5G-compliant waveform with a recording bandwidth of $61.44$ MHz.
\end{enumerate}

We emphasize that the generative processes of the signals above are not only considered unknown in the simulations, but are in fact truly unknown to the authors. %The generative processes of these signals are unknown in the simulations and to the authors.
The dataset examples for the first three types (EMISignal1, CommSignal2, and CommSignal3) were recorded over-the-air, while the last one (CommSignal5G1) was generated and recorded within a controlled wired laboratory environment, with wireless impairments introduced via simulators.

\begin{samepage}
\begin{figure*}[h!]   
\centering
  \begin{subfigure}{0.65\textwidth}
    \centering
    \includegraphics[width=\textwidth]{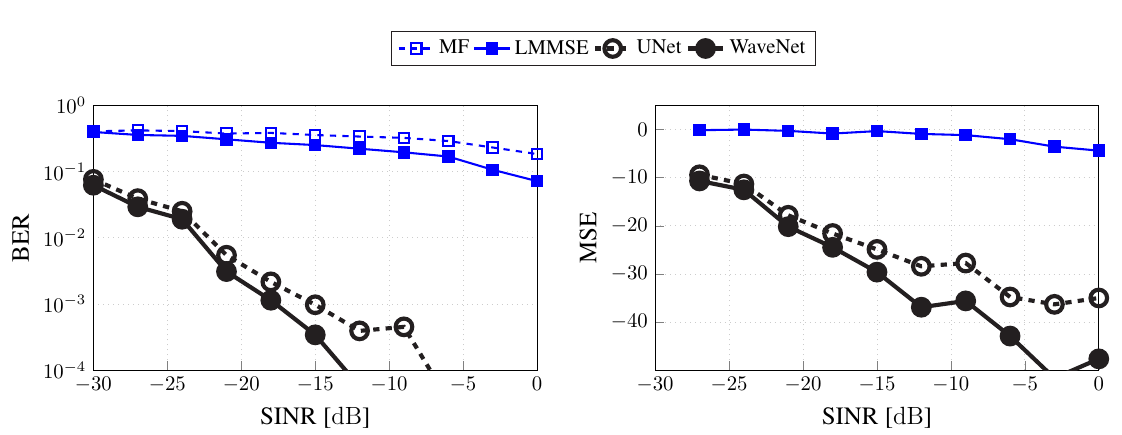} 
   \caption{\gls{qpsk} SOI.}
    \label{fig:ber_mse_qpsk_EMI}
  \end{subfigure}
  
  \vspace{-0.4cm}
  \begin{subfigure}{0.65\textwidth}
    \centering
    \includegraphics[width=\textwidth]{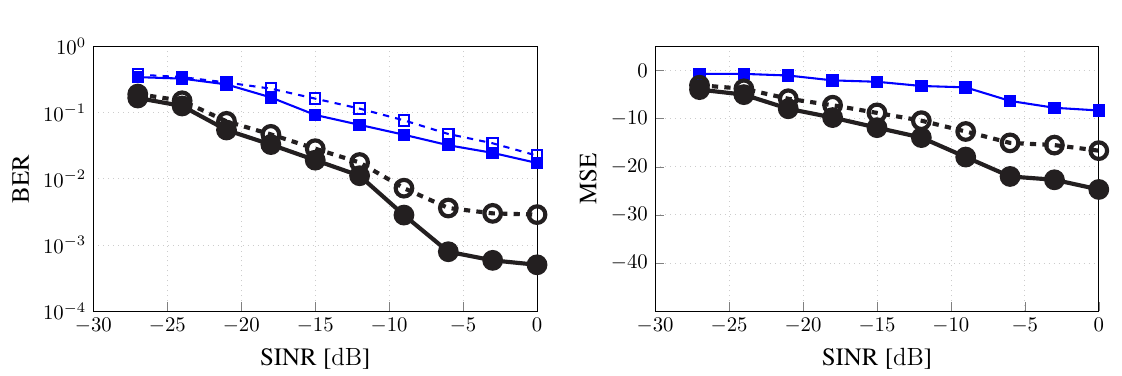} 
   \caption{OFDM-QPSK SOI.}
    \label{fig:ber_mse_ofdmqpsk_EMI}
  \end{subfigure}
  \caption{BER and MSE as a function of the target SINR for  {QPSK} and {OFDM-QPSK}
  \gls{soi}, with EMISignal1 interference. This figure shows the benchmarks discussed in this paper, along with our proposed data-driven methods.}
  \label{fig:ber_mse_EMISignal1}
\end{figure*}
\begin{figure*}[h!]  
\centering
  \begin{subfigure}{0.65\textwidth}
    \centering
    \includegraphics[width=\textwidth]{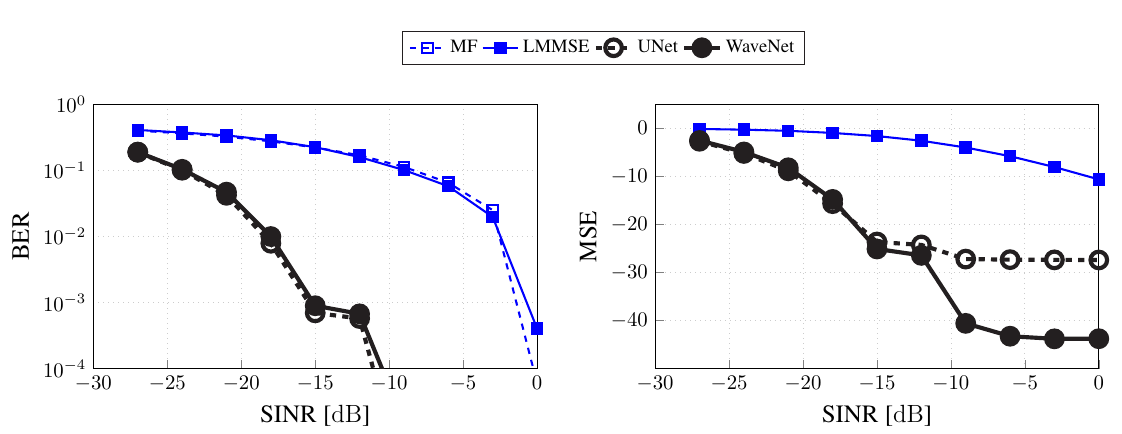} 
   \caption{QPSK SOI.}
    \label{fig:ber_mse_qpsk_Comm2}
  \end{subfigure}
  
  \vspace{-0.4cm}
  \begin{subfigure}{0.65\textwidth}
    \centering
    \includegraphics[width=\textwidth]{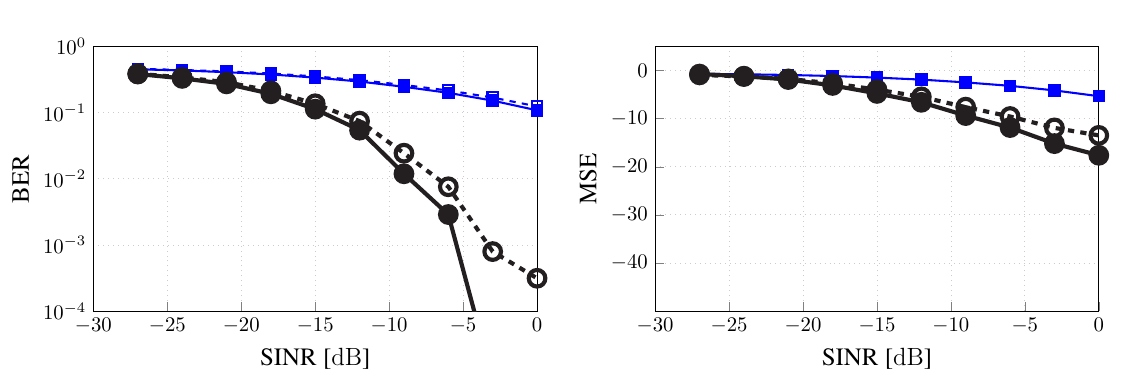} 
   \caption{OFDM-QPSK SOI.}
    \label{fig:ber_mse_ofdmqpsk_Comm2}
  \end{subfigure}
  \caption{BER and MSE as a function of the target SINR for {QPSK} and {OFDM-QPSK}
  \gls{soi}, with CommSignal2 interference. This figure shows the benchmarks discussed in this paper, along with our proposed data-driven methods.}
  \label{fig:ber_mse_commsignal2}
\end{figure*}
\end{samepage}

\begin{samepage}
\begin{figure*}[h!]  
\centering
  \begin{subfigure}{0.65\textwidth}
    \centering
    \includegraphics[width=\textwidth]{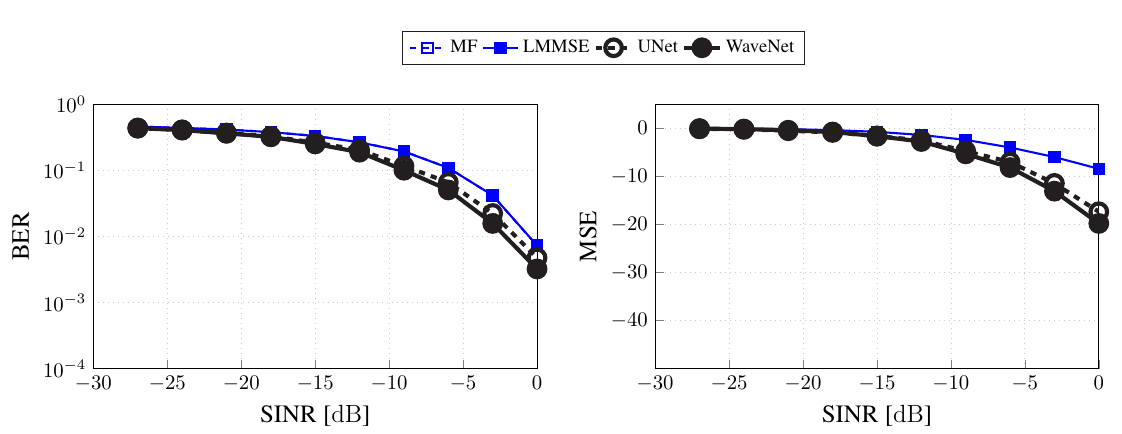} 
   \caption{QPSK SOI.}
    \label{fig:ber_mse_qpsk_Comm3}
  \end{subfigure}

  \vspace{-.4cm}
  \begin{subfigure}{0.65\textwidth}
    \centering
    \includegraphics[width=\textwidth]{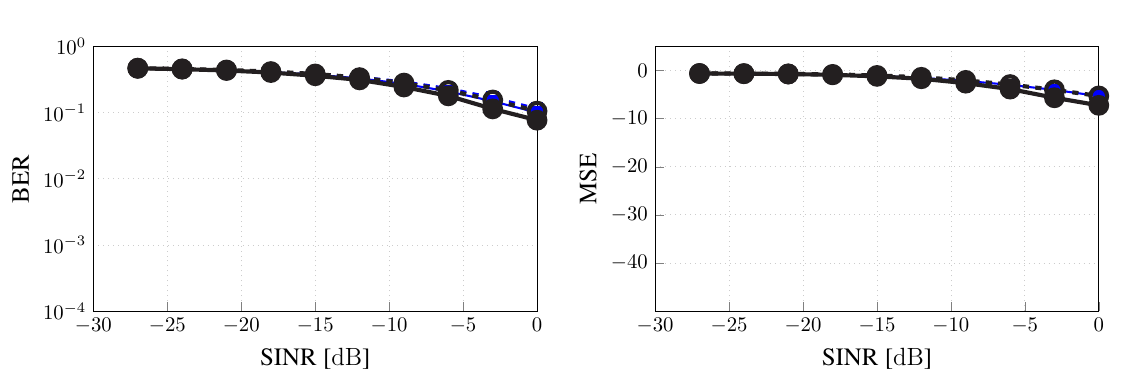} 
   \caption{OFDM-QPSK SOI.}
    \label{fig:ber_mse_ofdmqpsk_Comm3}
  \end{subfigure}
  \caption{BER and MSE as a function of the target SINR for {QPSK} and {OFDM-QPSK}
  \gls{soi}, with CommSignal3 interference. This figure shows the benchmarks discussed in this paper, along with our proposed data-driven methods.}
  \label{fig:ber_mse_qpsk}
\end{figure*}
\begin{figure*}[h!] 
\centering
  \begin{subfigure}{0.65\textwidth}
    \centering
    \includegraphics[width=\textwidth]{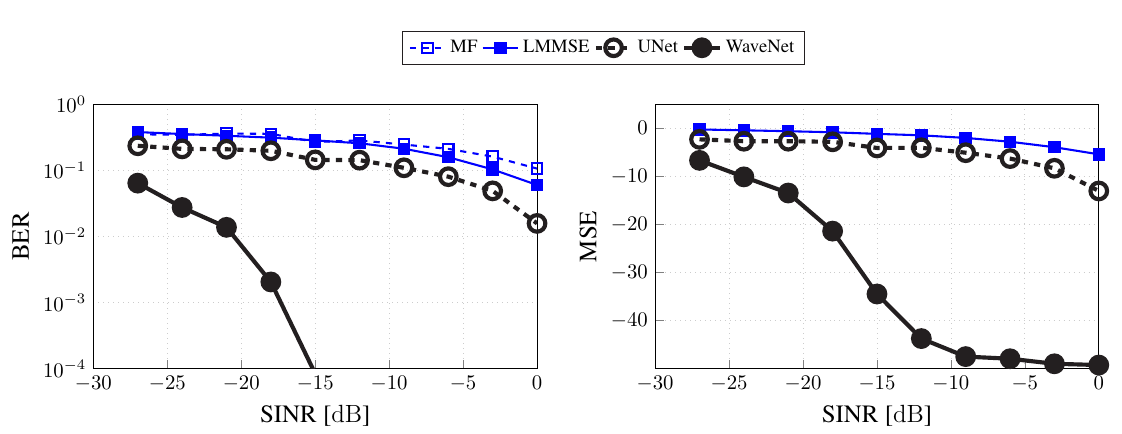} 
   \caption{QPSK SOI.}
    \label{fig:ber_mse_qpsk_Comm5G1}
  \end{subfigure}

  \vspace{-.4cm}
  \begin{subfigure}{0.65\textwidth}
    \centering
    \includegraphics[width=\textwidth]{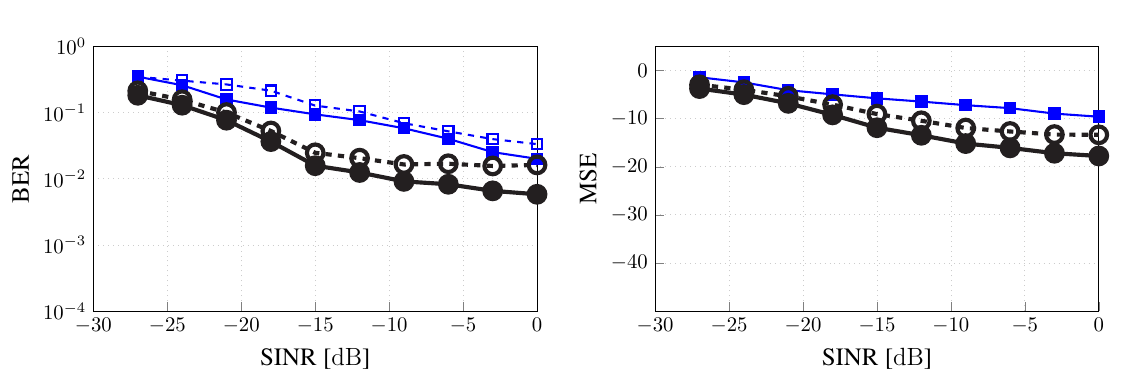} 
   \caption{OFDM-QPSK SOI.}
    \label{fig:ber_mse_ofdmqpsk_Comm5G1}
  \end{subfigure}
  \caption{BER and MSE as a function of the target SINR for {QPSK} and {OFDM-QPSK}  
\gls{soi}, with CommSignal5G1 interference. This figure shows the benchmarks discussed in this paper, along with our proposed data-driven methods.  
}
  \label{fig:ber_mse_commsignal5g1}
\end{figure*}
\end{samepage}

To create interference signal examples, we divided the set of examples into training and test sets. A frame of the respective interference type was then randomly selected (uniformly) from the corresponding set, and a random window of $N=40,960$ samples was extracted. Each interference component was scaled to achieve a target (empirical) \gls{sinr}. Since all signal datasets are normalized to have unit power, for a target \gls{sinr} level $\kappa^2=10^{(\textrm{\gls{sinr} in dB})/10}$, the interference signal is scaled by $1/\kappa$. Each interference frame $\rvecb^{(i)}$ also undergoes a random phase rotation before being added to the SOI $\rvecs^{(i)}$ to create a mixture example $\rvecy^{(i)}$ (see \eqref{eq:mixturemodel}). Note that we choose the term \gls{sinr} rather than SIR since some of the signals we use were recorded, thus they inevitably contain additive noise. Consequently, the effective interference in these cases is a sum of a non-Gaussian interference component and an additive noise component. %The scaling factor $\kappa$ and the instantaneous \gls{sinr} of each interference frame added to the mixture are available as metadata and could be used during training (but of course, not at inference time).

For the recorded interference  signals, the number of examples available per signal type changes. Furthermore, while the length of each recorded frame is the same for each signal type, it varies across types as well. In particular, we have $530$ examples of $230,000$ samples for EMISignal1, $100$ examples of $43,560$ for CommSignal2, $139$ examples of $260,000$ for CommSignal3, and $149$ examples of $230,000$ for CommSignal5G1. For consistency, we set the length of all input mixtures to $40,960$ samples. Note that our dataset consists of signals represented in baseband form, where the carrier frequency component was removed prior to saving. However, since the signals were not precisely centered around the presumed carrier frequency, their spectral content was not perfectly aligned around zero. This resulted in a non-zero DC level and only partial frequency overlap. To simulate the intended setting of maximal frequency overlap, we applied an additional frequency-shifting step to align the signals’ spectral content around zero, ensuring their proper alignment for the analysis presented in this work. In particular, signals EMISignal1 and CommSignal5G1 were shifted in frequency to have their spectral energy content lie in baseband frequencies, simulating co-channel interference that overlaps both in time and frequency. Figure~\ref{fig:dataset} shows the time- and frequency-domain representations of the recorded signal datasets used as interferences. Code examples can be found at \url{https://rfchallenge.mit.edu/icassp24-single-channel/}.

\subsection{OUR RESULTS}\label{sec:our_results}
Figures~\ref{fig:ber_mse_EMISignal1}--\ref{fig:ber_mse_commsignal5g1} show the performance of the two traditional interference rejection algorithms, introduced in Section~\ref{sec:trad_methods}, and our proposed deep learning-based interference rejection algorithms, introduced in Section~\ref{sec:data-driven}, over the eight possible \gls{soi}-interference combinations. The performance is measured in terms of \gls{ber} or \gls{mse} as a function of the target \gls{sinr}. The plots include the following curves:
\begin{itemize}
    \item \textbf{MF:} ignore the (potential) non-Gaussianity of the interference, and apply a matched filter to the mixture~\eqref{eq:mixturemodel}. We note that here \gls{mf} is applied only for the purpose of decoding (\gls{ber} plots only).
    \item \textbf{\gls{lmmse}:} the \gls{soi} is estimated via \gls{mse}-optimal linear estimation, the best-performing traditional method described in Section~\ref{sec:trad_methods}. Since the \gls{lmmse} requires the inversion of the covariance matrix $\matC_{\rvecy\rvecy}\in\complexset^{N\times N}$, which is generally of complexity $\setO(N^3)$, we have implemented this solution by applying \eqref{lmmsedef} to consecutive blocks of length $2,560$ samples each.\footnote{Computing the LMMSE for sequences of length $40,960$ samples is impractical, as it requires inverting a $40,960 \times 40,960$ matrix, leading to computations on the order of $40,960^3 \sim 10^{12}$. However, we computed the LMMSE using blocks of length $2,560$, which is already close to its asymptotic value and, in particular, to the LMMSE for sequences of length $40,960$ samples.}
    \item \textbf{UNet and WaveNet:} our proposed architectures, as presented in Section~\ref{sec:data-driven}. We emphasize that a separate neural network was trained for each mixture case.
    %\item \textbf{RF challenge submissions:} learning-based solutions from the research teams who submitted to the RF Challenge so far~\cite{Tian24_04,Damara24_04,TUB2024,LHen2024,IMEC2024}.\footnote{The RF challenge is hosted at \url{https://rfchallenge.mit.edu/}, where details on how to participate in the open challenges are provided.}
\end{itemize}

Both the LMMSE estimation approach and DNN-based interference rejection methods include a final step that treats residual interference as Gaussian, applying standard matched filtering prior to decoding to improve post-processing SNR. 

For the QPSK SOI case, this involves matched filtering, sampling at optimal points (assuming all necessary synchronization information is available), and then hard decoding to obtain symbols, which are mapped to bits. Similarly, for the OFDM SOI, we assume perfect synchronization, remove the cyclic prefix, apply an FFT of the appropriate size, and estimate the received symbols on active subcarriers, mapping them to their corresponding bits.

Both learning-based solutions described in Section~\ref{sec:data-driven} outperform the best traditional method out of the ones we consider as our benchmarks, namely \gls{lmmse} estimation, achieving up to two orders of magnitude performance improvements at considerably low \gls{sinr} values. For instance, in Fig.~\ref{fig:ber_mse_qpsk_Comm5G1}, at an \gls{sinr} level of $-18\,\mathrm{dB}$, the WaveNet model achieves a \gls{ber} of approximately $10^{-3}$ and an \gls{mse} below $-20\,\mathrm{dB}$. In contrast, the solutions based on \gls{lmmse} and no mitigation (other than MF) only reach a \gls{ber} slightly above $10^{-1}$ and an \gls{mse} around $0~\mathrm{dB}$.
  
\subsection{ICASSP'24 SP GRAND CHALLENGE RESULTS}

\begin{samepage}
\begin{figure*}[h!]  
\centering
  \begin{subfigure}{0.63\textwidth}
    \centering
    \includegraphics[width=\textwidth]{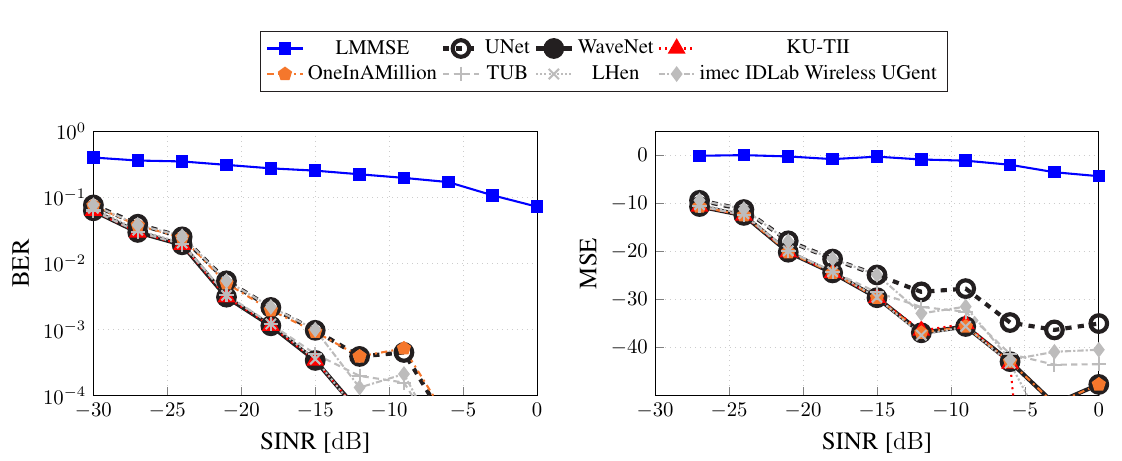} 
   \caption{QPSK SOI.}
    \label{fig:ber_mse_qpsk_EMI_icassp}
  \end{subfigure}

  \vspace{-.4cm}
  \begin{subfigure}{0.63\textwidth}
    \centering
    \includegraphics[width=\textwidth]{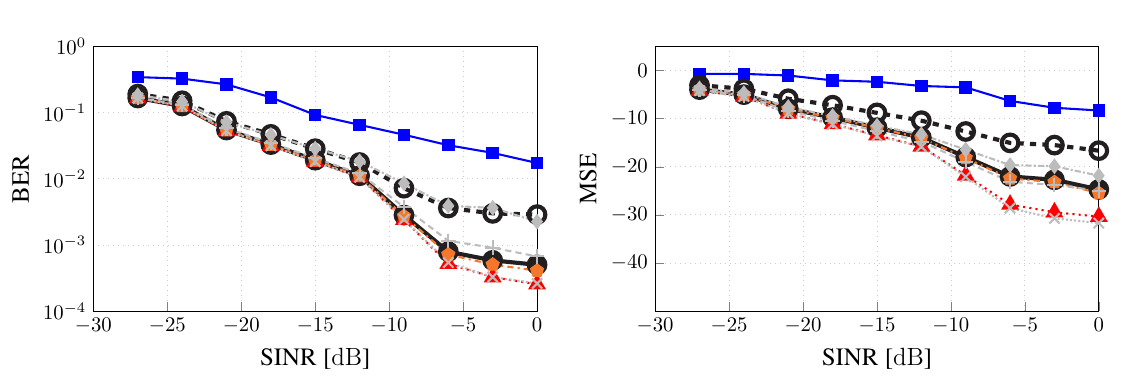} 
   \caption{OFDM-QPSK SOI.}
    \label{fig:ber_mse_ofdmqpsk_EMI_icassp}
  \end{subfigure}
  \caption{BER and MSE as a function of the target SINR for {QPSK} and {OFDM-QPSK}
  \gls{soi}, with EMISignal1 interference. This figure includes the LMMSE benchmark, along with our proposed data-driven methods and the proposed methods of the participants of the ICASSP'24 SP Grand Challenge.}
  \label{fig:ber_mse_EMISignal1_icassp}
\end{figure*}
\begin{figure*}[h!] 
\centering
  \begin{subfigure}{0.63\textwidth}
    \centering
    \includegraphics[width=\textwidth]{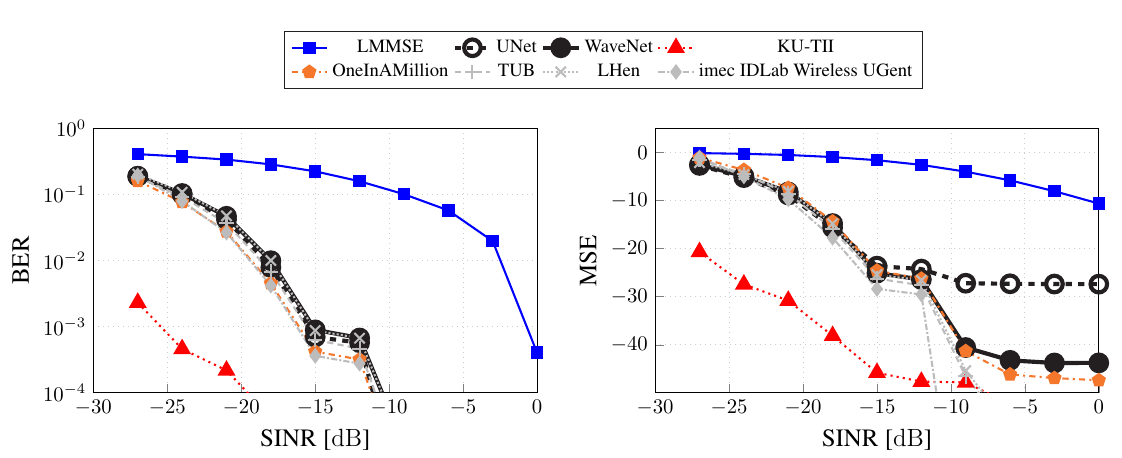} 
   \caption{QPSK SOI.}
    \label{fig:ber_mse_qpsk_Comm2_icassp}
  \end{subfigure}

  \vspace{-.4cm}
    \begin{subfigure}{0.63\textwidth}
    \centering
    \includegraphics[width=\textwidth]{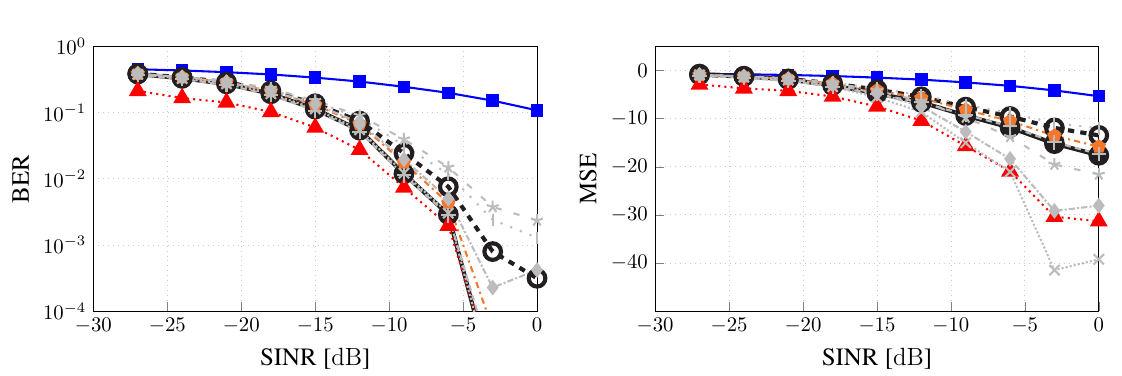} 
   \caption{OFDM-QPSK SOI.}
    \label{fig:ber_mse_ofdmqpsk_Comm2_icassp}
  \end{subfigure}
  \caption{BER and MSE as a function of the target SINR for {QPSK} and {OFDM-QPSK}
  \gls{soi}, with CommSignal2 interference. This figure includes the LMMSE benchmark, along with our proposed data-driven methods and the proposed methods of the participants of the ICASSP'24 SP Grand Challenge.}
  \label{fig:ber_mse_commsignal2_icassp}
\end{figure*}
\end{samepage}

\begin{samepage}
\begin{figure*}[h!] 
\centering
  \begin{subfigure}{0.63\textwidth}
    \centering
    \includegraphics[width=\textwidth]{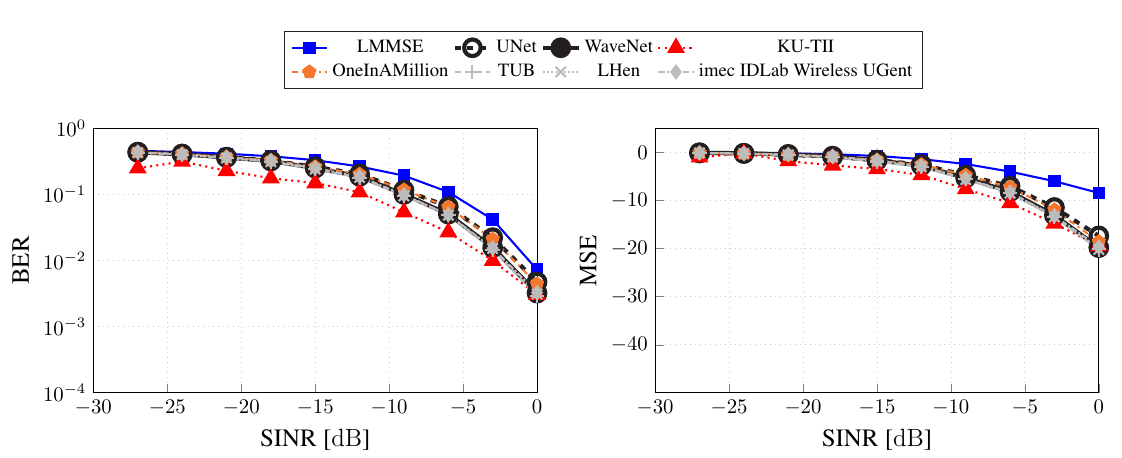} 
   \caption{QPSK SOI.}
    \label{fig:ber_mse_qpsk_Comm3_icassp}
  \end{subfigure}

  \vspace{-.4cm}
\begin{subfigure}{0.63\textwidth}
    \centering
    \includegraphics[width=\textwidth]{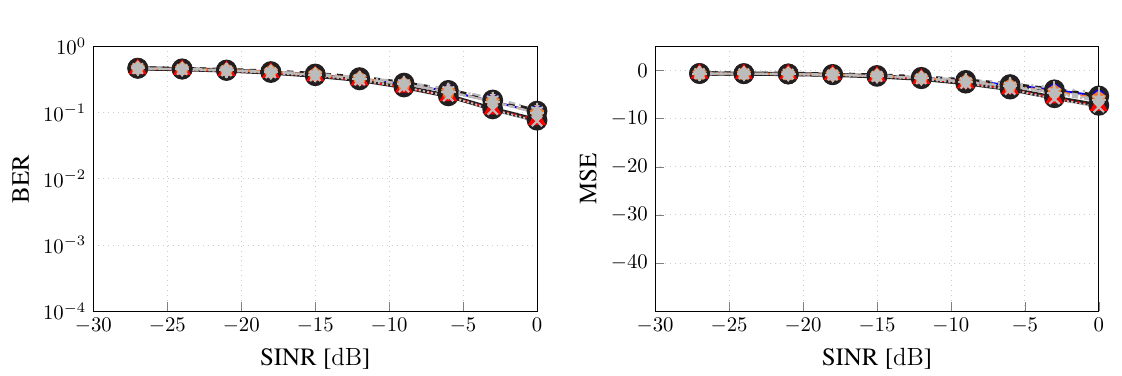} 
   \caption{OFDM-QPSK SOI.}
    \label{fig:ber_mse_ofdmqpsk_Comm3_icassp}
  \end{subfigure}
  \caption{BER and MSE as a function of the target SINR for {QPSK} and {OFDM-QPSK}
  \gls{soi}, with CommSignal3 interference. This figure includes the LMMSE benchmark, along with our proposed data-driven methods and the proposed methods of the participants of the ICASSP'24 SP Grand Challenge.}
  \label{fig:ber_mse_commsignal3_icassp}
\end{figure*}
\begin{figure*}[h!] 
\centering
  \begin{subfigure}{0.63\textwidth}
    \centering
    \includegraphics[width=\textwidth]{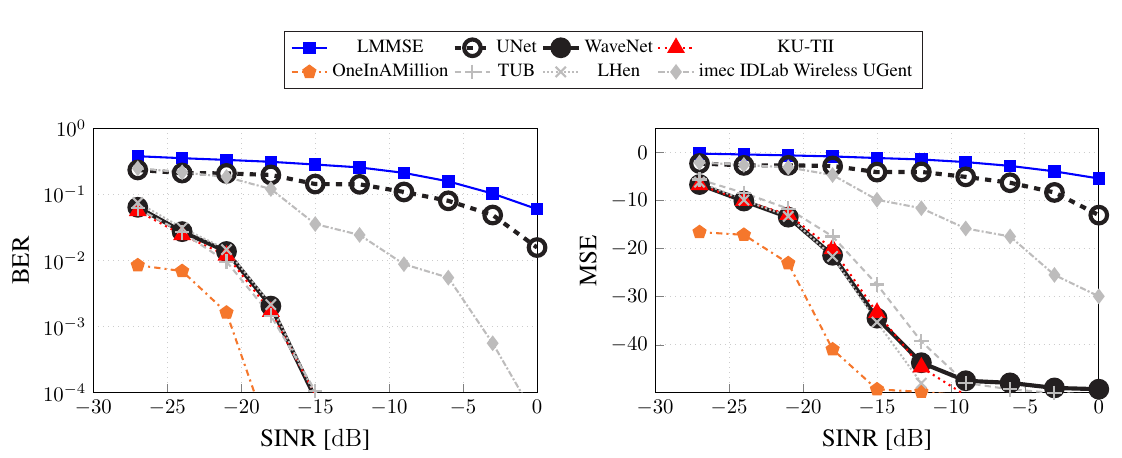} 
   \caption{CommSignal5G1 interference.}
    \label{fig:ber_mse_qpsk_Comm5G1_icassp}
  \end{subfigure}

  \vspace{-.4cm}
  \begin{subfigure}{0.63\textwidth}
    \centering
    \includegraphics[width=\textwidth]{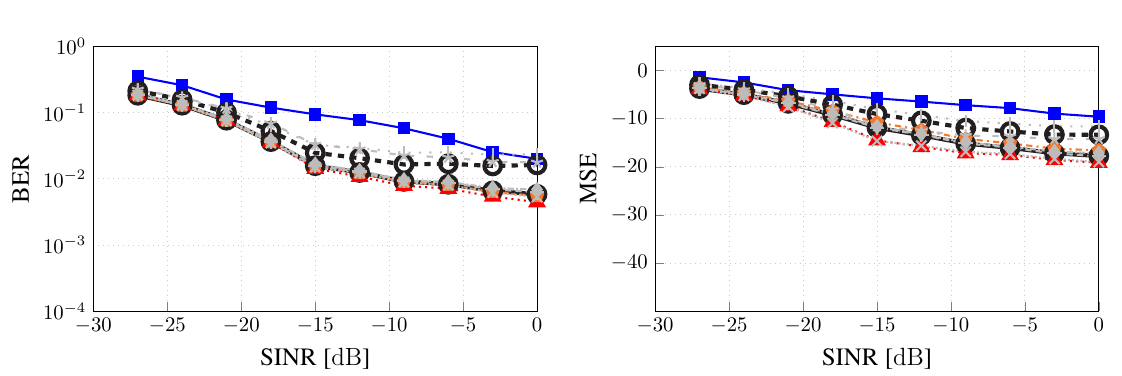} 
   \caption{CommSignal5G1 interference.}
    \label{fig:ber_mse_ofdmqpsk_Comm5G1_icassp}
  \end{subfigure}
  \caption{BER and MSE as a function of the target SINR for {QPSK} and {OFDM-QPSK}
  \gls{soi}, with CommSignal5G1 interference. This figure includes the LMMSE benchmark, along with our proposed data-driven methods and the proposed methods of the participants of the ICASSP'24 SP Grand Challenge.}
  \label{fig:ber_mse_commsignal5g1_icassp}
\end{figure*}
\end{samepage}
We hosted the ``Data-Driven Radio Frequency Signal Separation Challenge" as part of the ICASSP'24 Signal Processing Grand Challenges~\cite{datadrivenrf2024}. Among the submissions, only a few significantly outperformed the learning-based benchmark methods described in Section~\ref{sec:data-driven} at specific mixture cases \cite{Tian24_04,Damara24_04}. The proposed solutions are summarized below:

\begin{itemize}
    \item KU-TII \cite{Tian24_04}: This team enhanced the WaveNet architecture provided as a benchmark. Their main contributions were two-fold: (i) Model-related: introducing learnable dilated convolutions within the convolutional layer; (ii) Data augmentation: expanding the training set by using as a validation set the test example data provided by the challenge organizers, and generating additional training examples for CommSignal2 by converting high-SNR waveforms with probably zero \gls{ber} into bits, then reconverting these bits back into waveforms to subtract them from the mixtures. With this data augmentation strategy, the team trained their algorithm with more data, potentially resulting in further performance gains unrelated to architectural improvements.\footnote{Unfortunately, in the ICASSP'24 Grand Challenge, the SOIs were accidentally identical between test and validation sets. The KU-TII team's training method for CommSignal2 interference only reused SOIs from the validation set, which allowed for test-set leakage. This may explain the outlier performance on CommSignal2.}

    \item One-In-A-Million \cite{Damara24_04}: This team presented two approaches, a Transformer UNet and a finetuned discriminative WaveNet. The Transformer UNet is a convolution-attention-based model with an encoder-decoder architecture that includes self-attention blocks in the bottleneck to refine representations, similar to the one introduced in~\cite{kong22}. Between these two options, the finetuned WaveNet achieved superior performance across all signal mixture cases, suggesting that Transformer architectures may require specific design modifications to fully leverage their capabilities in the source separation of digital communication signals.

    \item LHen \cite{LHen2024}: This team extended the WaveNet baseline by incorporating an autoencoder tailored to the modulation type of the \gls{soi}. The encoder learns to demodulate the waveform estimated by WaveNet, while the corresponding decoder re-synthesizes the SOI waveform from the extracted bit sequence to achieve low MSE.
    
    \item TUB \cite{TUB2024}: This team employed a DEMUCS architecture, featuring an encoder-decoder framework with a convolutional encoder, bidirectional LSTM applied on the encoder output, and a convolutional decoder, all linked via UNet skip connections \cite{defossez20,defossez21}. They further adapted DEMUCS for direct bit regression, similar to the ``Bit Regression" baseline from the ``Single-Channel RF Challenge" hosted on the RF Challenge website in 2021.\footnote{The bit regression baseline code is available at \url{https://github.com/RFChallenge/rfchallenge_singlechannel_starter/tree/main/example/demod_bitregression}.}
    
    \item imec-IDLab \cite{IMEC2024}: This team developed a UNet architecture, building upon the baseline UNet provided in the challenge, redesigned specifically for separating interference signals in the time-frequency domain. For example, they leveraged domain-specific knowledge of \glspl{soi} for the OFDM-\gls{qpsk} \gls{soi} case by integrating elements of OFDM signal resource grid configurations, such as the cyclic prefix, into the architecture responsible for the decoding process. 
\end{itemize}

Similar to the previous section, Figures~\ref{fig:ber_mse_EMISignal1_icassp}--\ref{fig:ber_mse_commsignal5g1_icassp} show the performance of the traditional interference rejection algorithm based on \gls{lmmse} estimation, our proposed deep learning-based interference rejection algorithms introduced in Section~\ref{sec:data-driven}, and the top-5-performing teams that participated in the challenge. 

As we can see, there are some teams whose solutions improved upon the baselines in some cases involving EMISignal1, CommSignal2, and CommSignal5G1 (see Figs.~\ref{fig:ber_mse_ofdmqpsk_EMI_icassp}, \ref{fig:ber_mse_commsignal2_icassp}, and  \ref{fig:ber_mse_qpsk_Comm5G1_icassp}). For example, ``KU-TII''~\cite{Tian24_04} especially shines in those mixture involving CommSignal2 interference, where they gain more than an order of magnitude in \gls{ber} at \gls{sinr} values below $-20\,{\rm dB}$ compared to our baseline architectures, and ``OneInAMillion''~\cite{Damara24_04} performs especially well in the \gls{qpsk} + CommSignal5G mixture, where they almost achieved an order of magnitude gain in \gls{ber} at \gls{sinr} values below $-20\,{\rm dB}$. Conversely, mixtures with CommSignal3 consistently challenge all methods. While we believe it is a multicarrier signal with a high data rate, the specific reasons for the difficutly to separate CommSignal3 from the \gls{soi} remain unclear, warranting further investigation. We note that, at least from a theoretical perspective, it should come as no surprise that one architecture is superior to others with respect to one fidelity measure (e.g., \gls{mse}), and is no longer superior with respect to a different one (e.g., minimum error probability). Indeed, a \gls{dnn} is trained for a specific goal, via choosing a single objective function, and therefore cannot necessarily be optimal in more than one sense.\footnote{However, this may happen in special cases, such as the \gls{mmse} and \gls{map} estimators in the Gaussian signal model.}

These results show the potential of data-driven, deep-learning-based solutions to provide significant improvements in interference rejection tasks when the interference has unknown structures that can be learned. They also demonstrate that innovative solutions are needed to achieve further performance gains for challenging signals such as CommSignal3.

\section{FUTURE DIRECTIONS AND CONCLUDING REMARKS}\label{sec:outlook}

In this section, we explore potential future research directions that can further advance the field of data-driven source separation of \gls{rf} signals using deep learning techniques. We end the section with concluding remarks, encapsulating the key findings of this paper.

\subsection{FUTURE RESEARCH DIRECTIONS}
\label{sec:future}
As demonstrated in the previous section---for a variety of signals using several methods---data-driven deep learning algorithms for \gls{rf} source separation can yield significant performance gains. However, to make these techniques practically relevant,  many aspects of this problem remain to be explored. Below, we outline and briefly discuss some of the more important ones.

% \paragraph*{SOIs with Unknown Generation Process}
%
\begin{figure}[t!]     
  \begin{subfigure}{0.8\columnwidth}
    \centering
    \includegraphics[width=\textwidth]{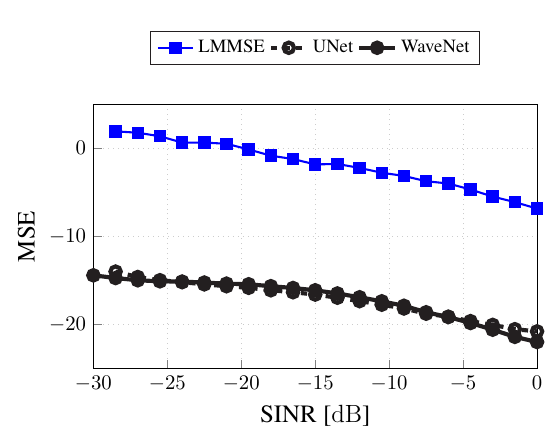} 
   \caption{EMISignal1 interference.}
    \label{fig:mse_Comm2_EMI}
  \end{subfigure}
  
  \begin{subfigure}{0.8\columnwidth}
    \centering
    \includegraphics[width=\textwidth]{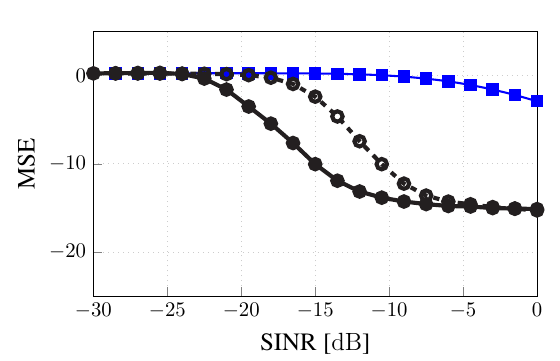} 
   \caption{CommSignal3 interference.}
    \label{fig:mse_Comm2_Comm3}
  \end{subfigure}

  \begin{subfigure}{0.8\columnwidth}
    \centering
    \includegraphics[width=\textwidth]{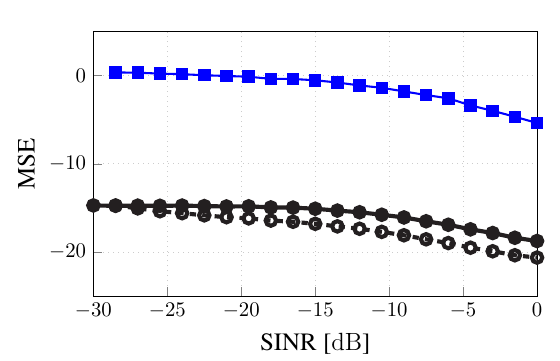} 
   \caption{CommSignal5G1 interference.}
    \label{fig:mse_Comm2_Comm5G1}
  \end{subfigure}
  \caption{MSE as a function of the target SINR for all combinations of CommSignal2 \gls{soi} and the rest of interference types considered in this work.}
  \label{fig:mse_Comm2}
\end{figure}
One natural extension of this work is to investigate scenarios where the generation process of the \gls{soi} is \emph{unknown}. Such cases are more challenging, as there is less prior information to exploit when designing the overall separation algorithm. For example, the \gls{mse} results obtained in Section~\ref{sec:results_SOI_knwon} can potentially be further improved if we leverage knowledge of the modulation scheme of the \gls{soi} via the decoding processing chain subsequent to the DNN-based interference mitigation module. Clearly, when the signal generation process is unknown, this approach is not applicable, and more sophisticated \gls{dnn} architectures would be required to achieve the same performance. 

As a preliminary empirical study on this front, we have  used CommSignal2 as the \gls{soi}, and evaluated the \gls{mse} using the UNet and WaveNet architectures (Section~\ref{sec:methods}), as well as LMMSE (Section~\ref{sec:trad_methods}). We used the remaining signals in the dataset as interference signals. The results, presented in Fig.~\ref{fig:mse_Comm2}, demonstrate that both the UNet and the WaveNet architectures outperforms LMMSE, except at low \gls{sinr} levels with CommSignal3 as the interference.

Yet another extension to a prevalent setting is to \gls{mimo} systems, which are a cornerstone of modern wireless communication standards such as 5G and Wi-Fi. In this case, nontrivial spatial patterns can potentially be learned and exploited. Although the multi-channel source separation literature is significantly richer than the single-channel one (e.g.,~\cite{hyvarinen1999fast,belouchrani1997blind,yeredor2002non,weiss2019maximum}), the discussion revolving \emph{data-driven deep-learning-based} methods in this context has not yet been comparatively addressed. Exploring how to utilize such spatial patterns \emph{in tandem} with the signals' unique statistical characteristics, including temporal structures, constitutes a promising avenue for future research.

Other important extensions refer to potential physical effects induced by the channel. For example,  a more comprehensive signal model may incorporate effects such as arbitrary time shifts and fading. More generally, an extended model of \eqref{eq:mixturemodel} can be expressed as:

\begin{equation}
    \rndy[n] = \setH\left\{\rnds[n]\right\} + \rndb[n-\rndk_b], \,\, n\in\integers, \label{eq:proc_model}
\end{equation}
where, as before, $\rnds[n]$ and $\rndb[n]$ are the \gls{soi} and interference, respectively, and $\setH\{\cdot\}$ denotes the channel response. For example, the channel could be $\setH\left\{\rnds[n]\right\}=\alpha[n]\cdot\rnds[n-\rndk_s]$ for some time-varying fading coefficient $\alpha[n]\in\complexset$ and an unknow delay $\rndk_s\in\naturals$ (e.g., \cite{lancho2022data}). In particular, $\setH\{\cdot\}$ is not necessarily linear or time-invariant. The advantage of the data-driven approach is that it can potentially learn to compensate for nonlinear, time-varying effects, provided they are governed by some learnable statistical law, and are well captured in the available datasets for training.

A central attribute of the data-driven solution approach presented in this work is that each interference mitigation module, once trained, is tailored for specific mixtures of \gls{soi} and interference types. While this approach can lead to statistically optimal separation performance, its robustness is not guaranteed. Furthermore, it is highly demanding in terms of the system resources, such as memory, to maintain a separate NN for each mixture type, which can be impractical in many scenarios. A desirable, more efficient, robust alternative would be to maintain a \emph{single} model capable of mitigating \emph{multiple} interference types. Such a model can establish the groundwork for developing a foundational RF signal separation model~\cite{bommasani2021opportunities,fontaine2024}, where signal characteristics (e.g., codes, modulations, pulse shaping) act as the ``modalities".

Recent work~\cite{alikhani24} introduced a foundation model for wireless channels, using an encoder-decoder transformer trained to predict masked channel embeddings, successfully addressing downstream tasks like beam prediction and LOS/NLOS classification. Similarly, exploring a foundation model trained for interference rejection could unlock downstream tasks such as demodulation or constellation classification. A crucial first step in this direction is demonstrating that a single DNN can handle multiple interference types with performance comparable to specialized models, in terms of both inference time and separation fidelity.

\subsection{CONCLUDING REMARKS}

In this paper we illustrate the potential of deep learning-based methods for source separation of \gls{rf} signals. Specifically, we show that mitigating strong unintentional interference from other \gls{rf} emitting sources operating at the same time and the same frequency band with data-driven methods leads to considerable gains relative to traditional, ``hand-crafted" methods. Through extensive simulation experiments, we demonstrate the superior performance of deep learning architectures, such as UNet and WaveNet, over the traditional signal processing methods of matched filtering and LMMSE estimation across various scenarios. Results from different leading research teams that participated in the  ``Data-Driven Radio Frequency Signal Separation Challenge", hosted as an ICASSP'24 Signal Processing Grand Challenge~\cite{datadrivenrf2024}, show that further improving significantly beyond the established deep-learning benchmarks is nontrivial, especially in mixtures involving multi-carrier signals.

Ultimately, these results represent merely an initial phase of a more extensive journey towards integrating \gls{ai} capabilities into receivers for enhanced interference rejection. The path forward would involve addressing additional, theoretical and practical, related problems, including interpretability of \glspl{dnn}, and presenting viable solutions to demonstrate the tangible benefits of these approaches.  Indeed, the results motivate further research and development for this dynamic domain within the broader community, with the ultimate goal of significantly improving future generations of \gls{rf} systems spanning diverse applications.

\bibliographystyle{IEEEbib}

\begin{thebibliography}{10}

  \bibitem{qualcomm_AR}
  Qualcomm~Technologies Inc.,
  \newblock ``{VR} and {AR} pushing connectivity limits,''
    \url{https://www.qualcomm.com/content/dam/qcomm-martech/dm-assets/documents/presentation_-_vr_and_ar_are_pushing_connectivity_limits_-web_0.pdf},
  \newblock Accessed: 2024-02-26.
  
  \bibitem{Hirzallah17}
  Mohammed Hirzallah, Wessam Afifi, and Marwan Krunz,
  \newblock ``Full-duplex-based rate/mode adaptation strategies for
    {W}i-{F}i/{LTE-U} coexistence: {A POMDP} approach,''
  \newblock {\em {IEEE} J. Sel. Areas Commun.}, vol. 35, no. 1, pp. 20--29, Nov.
    2017.
  
  \bibitem{Naik21}
  Gaurang Naik, Jung-Min Park, Jonathan Ashdown, and William Lehr,
  \newblock ``Next generation {W}i-{F}i and {5G NR-U} in the 6 {GHz} bands:
    {O}pportunities and challenges,''
  \newblock {\em IEEE Access}, vol. 8, pp. 153027--153056, Aug. 2020.
  
  \bibitem{oyedare2022interference}
  Taiwo Oyedare, Vijay~K. Shah, Daniel~J. Jakubisin, and Jeff~H. Reed,
  \newblock ``Interference suppression using deep learning: Current approaches
    and open challenges,''
  \newblock {\em IEEE Access}, June 2022.
  
  \bibitem{vantrees_01}
  Harry L.~Van Trees,
  \newblock {\em Detection, Estimation, and Modulation Theory, Part I:},
  \newblock Wiley, New York, NY, USA, 2001.
  
  \bibitem{tu2007particle}
  Tu~Shilong, Chen Shaohe, Zheng Hui, and Wan Jian,
  \newblock ``Particle filtering based single-channel blind separation of
    co-frequency {MPSK} signals,''
  \newblock in {\em IEEE Int. Symp. Intell. Signal Process. and Commun. Syst.},
    Feb. 2007, pp. 582--585.
  
  \bibitem{tu2008single}
  Tu~Shilong, Zheng Hui, and Gu~Na,
  \newblock ``Single-channel blind separation of two {QPSK} signals using
    per-survivor processing,''
  \newblock in {\em IEEE Asia Pac. Conf. Circuits Syst. (APCCAS)}, Dec. 2008, pp.
    473--476.
  
  \bibitem{lee2011interference}
  Jungwon Lee, Dimitris Toumpakaris, and Wei Yu,
  \newblock ``Interference mitigation via joint detection,''
  \newblock {\em {IEEE} J. Sel. Areas Commun.}, vol. 29, no. 6, pp. 1172--1184,
    2011.
  
  \bibitem{chevalier2018third}
  Pascal Chevalier, Jean-Pierre Delmas, and Mustapha Sadok,
  \newblock ``Third-order {V}olterra {MVDR} beamforming for non-{G}aussian and
    potentially non-circular interference cancellation,''
  \newblock {\em {IEEE} Trans. Signal Process.}, vol. 66, no. 18, pp. 4766--4781,
    July 2018.
  
  \bibitem{o2016radio}
  Timothy~J. O'shea and Nathan West,
  \newblock ``Radio machine learning dataset generation with {GNU} radio,''
  \newblock in {\em Proc. GNU Radio Conf.}, Sept. 2016, vol.~1.
  
  \bibitem{rfchallenge}
  {MIT RLE},
  \newblock ``{RF Challenge - AI Accelerator},''
    \url{https://rfchallenge.mit.edu},
  \newblock Accessed 2024-12-03.
  
  \bibitem{stoller2018wave}
  Daniel Stoller, Sebastian Ewert, and Simon Dixon,
  \newblock ``{Wave-U-Net}: A multi-scale neural network for end-to-end audio
    source separation,''
  \newblock in {\em Intl. Soc. for Music Inf. Retrieval Conf.}, May 2018, pp.
    334--340.
  
  \bibitem{nugraha2016multichannel}
  Aditya~Arie Nugraha, Antoine Liutkus, and Emmanuel Vincent,
  \newblock ``Multichannel audio source separation with deep neural networks,''
  \newblock {\em IEEE/ACM Trans. Audio, Speech, Lang. Process.}, vol. 24, no. 9,
    pp. 1652--1664, June 2016.
  
  \bibitem{gandelsman2019double}
  Yosef Gandelsman, Assaf Shocher, and Michal Irani,
  \newblock ``{``Double-DIP"}: Unsupervised image decomposition via coupled
    deep-image-priors,''
  \newblock in {\em Proc. of IEEE/CVF Conf. Comput. Vis. Pattern Recognit.
    (CVPR)}, June 2019, pp. 11026--11035.
  
  \bibitem{huang2015joint}
  Po-Sen Huang, Minje Kim, Mark Hasegawa-Johnson, and Paris Smaragdis,
  \newblock ``Joint optimization of masks and deep recurrent neural networks for
    monaural source separation,''
  \newblock {\em IEEE/ACM Trans. Audio, Speech, Lang. Process.}, vol. 23, no. 12,
    pp. 2136--2147, Dec. 2015.
  
  \bibitem{jayashankar2023score-neurips}
  Tejas Jayashankar, Gary~C.F. Lee, Alejandro Lancho, Amir Weiss, Yury
    Polyanskiy, and Gregory~W. Wornell,
  \newblock ``Score-based source separation with applications to digital
    communication signals,''
  \newblock in {\em Advances Neural Inform.\ Proc.\ Syst.\ (NeurIPS)}, New
    Orleans, LA, Dec. 2023.
  
  \bibitem{lee2023neural}
  Gary~C.F. Lee, Amir Weiss, Alejandro Lancho, Yury Polyanskiy, and Gregory~W.
    Wornell,
  \newblock ``On neural architectures for deep learning-based source separation
    of co-channel {OFDM} signals,''
  \newblock in {\em IEEE Int. Conf. Acoust. Speech Signal Process.}, June 2023.
  
  \bibitem{github1}
  {MIT RLE: RF Challenge},
  \newblock ``Single-channel source separation: Preliminary test of neural
    network architectures,''
    \url{https://github.com/RFChallenge/SCSS_DNN_Comparison}, 2022.
  
  \bibitem{oyedare2024comprehensive}
  Taiwo~Remilekun Oyedare,
  \newblock {\em A Comprehensive Analysis of Deep Learning for Interference
    Suppression, Sample and Model Complexity in Wireless Systems},
  \newblock Phd thesis, Virginia Tech, Mar. 2024.
  
  \bibitem{rfdeepai}
  {DeepSig Inc.},
  \newblock ``{RF Datasets For Machine Learning},''
    \url{https://www.deepsig.ai/datasets},
  \newblock Accessed 2024-12-03.
  
  \bibitem{iqengine}
  {IQ Engine},
  \newblock ``{A web-based SDR toolkit for analyzing, processing, and sharing RF
    recordings},'' \url{https://iqengine.org/},
  \newblock Accessed 2024-12-03.
  
  \bibitem{lee2022exploiting}
  Gary~C.F. Lee, Amir Weiss, Alejandro Lancho, Jennifer Tang, Yuheng Bu, Yury
    Polyanskiy, and Gregory~W. Wornell,
  \newblock ``Exploiting temporal structures of cyclostationary signals for
    data-driven single-channel source separation,''
  \newblock in {\em IEEE Int. Workshop Mach. Learn. Signal Process. (MLSP)}, Aug.
    2022.
  
  \bibitem{lecun1998gradient}
  Yann LeCun, L{\'e}on Bottou, Yoshua Bengio, and Patrick Haffner,
  \newblock ``Gradient-based learning applied to document recognition,''
  \newblock {\em Proc. IEEE}, vol. 86, no. 11, pp. 2278--2324, Aug. 1998.
  
  \bibitem{deng2009imagenet}
  Jia Deng, Wei Dong, Richard Socher, Li-Jia Li, Kai Li, and Li~Fei-Fei,
  \newblock ``{I}magenet: A large-scale hierarchical image database,''
  \newblock in {\em IEEE Conf. Comput. Vis. Pattern Recognit. ({CVPR})}, Aug.
    2009, pp. 248--255.
  
  \bibitem{cook2014vast}
  Kristin Cook, Georges Grinstein, and Mark Whiting,
  \newblock ``The {VAST} challenge: History, scope, and outcomes: An introduction
    to the special issue,'' 2014.
  
  \bibitem{luszczek2005introduction}
  Piotr Luszczek, Jack~J. Dongarra, David Koester, Rolf Rabenseifner, Bob Lucas,
    Jeremy Kepner, John McCalpin, David Bailey, and Daisuke Takahashi,
  \newblock ``Introduction to the {HPC} challenge benchmark suite,''
  \newblock {\em Lawrence Berkeley Nat. Lab., Berkeley, CA}, 2005.
  
  \bibitem{datadrivenrf2024}
  Tejas Jayashankar, Benoy Kurien, Alejandro Lancho, Gary~C.F. Lee, Yury
    Polyanskiy, Amir Weiss, and Gregory Wornell,
  \newblock ``The data-driven radio frequency signal separation challenge,''
  \newblock in {\em Proc. IEEE Int. Conf. Acoust., Speech, Signal Process.
    (ICASSP)}, Apr. 2024.
  
  \bibitem{hwang2008ofdm}
  Taewon Hwang, Chenyang Yang, Gang Wu, Shaoqian Li, and Geoffrey~Ye Li,
  \newblock ``{OFDM} and its wireless applications: {A} survey,''
  \newblock {\em {IEEE} Trans. Veh. Technol.}, vol. 58, no. 4, pp. 1673--1694,
    Aug. 2008.
  
  \bibitem{ronneberger2015unet}
  Olaf Ronneberger, Philipp Fischer, and Thomas Brox,
  \newblock ``{U-Net}: Convolutional networks for biomedical image
    segmentation,''
  \newblock in {\em Med. Image Comput. Comput. Assist. Interv.} Nov. 2015, pp.
    234--241, Springer.
  
  \bibitem{akeret2017radio}
  Joel Akeret, Chihway Chang, Aurelien Lucchi, and Alexandre Refregier,
  \newblock ``Radio frequency interference mitigation using deep convolutional
    neural networks,''
  \newblock {\em Astronomy and Computing}, vol. 18, pp. 35--39, Jan. 2017.
  
  \bibitem{tzinis2020sudo}
  Efthymios Tzinis, Zhepei Wang, and Paris Smaragdis,
  \newblock ``Sudo {RM}-{RF}: Efficient networks for universal audio source
    separation,''
  \newblock in {\em Proc. IEEE Int. Workshop Mach. Learn. Signal Process.
    (MLSP)}, Sept. 2020.
  
  \bibitem{picinbono1995widely}
  Bernard Picinbono and Pascal Chevalier,
  \newblock ``Widely linear estimation with complex data,''
  \newblock {\em IEEE Trans. Signal Process.}, vol. 43, no. 8, pp. 2030--2033,
    Aug. 1995.
  
  \bibitem{Oord16}
  A{\"{a}}ron van~den Oord, Sander Dieleman, Heiga Zen, Karen Simonyan, Oriol
    Vinyals, Alex Graves, Nal Kalchbrenner, Andrew~W. Senior, and Koray
    Kavukcuoglu,
  \newblock ``Wavenet: {A} generative model for raw audio,''
  \newblock {\em ar{X}iv:1609.03499}, vol. abs/1609.03499, Sept. 2016.
  
  \bibitem{Rethage18}
  Dario Rethage, Jordi Pons, and Xavier Serra,
  \newblock ``A wavenet for speech denoising,''
  \newblock in {\em Proc. IEEE Int. Conf. Acoust., Speech, Signal Process.
    (ICASSP)}, Apr. 2018, pp. 5069--5073.
  
  \bibitem{Tian24_04}
  Yu~Tian, Ahmed Alhammadi, Abdullah Quran, and Abubakar~Sani Ali,
  \newblock ``A novel approach to wavenet architecture for {RF} signal separation
    with learnable dilation and data augmentation,''
  \newblock in {\em Proc. IEEE Int. Conf. Acoust., Speech, Signal Process.
    Workshops (ICASSPW)}, Apr. 2024, pp. 79--80.
  
  \bibitem{Damara24_04}
  Fadli Damara, Zoran Utkovski, and Slawomir Stanczak,
  \newblock ``Signal separation in radio spectrum using self-attention
    mechanism,''
  \newblock in {\em Proc. IEEE Int. Conf. Acoust., Speech, Signal Process.
    Workshops (ICASSPW)}, Apr. 2024, pp. 99--100.
  
  \bibitem{kong22}
  Zhifeng Kong, Wei Ping, Ambrish Dantrey, and Bryan Catanzaro,
  \newblock ``Speech denoising in the waveform domain with self-attention,''
  \newblock in {\em Proc. IEEE Int. Conf. Acoust., Speech, Signal Process.
    (ICASSP)}, May 2022, pp. 7867--7871.
  
  \bibitem{LHen2024}
  Lukas Henneke,
  \newblock ``Improving data-driven {RF} signal separation with {SOI}-matched
    autoencoders,''
  \newblock in {\em Proc. IEEE Int. Conf. Acoust., Speech, Signal Process.
    Workshops (ICASSPW)}, Apr. 2024, pp. 45--46.
  
  \bibitem{TUB2024}
  Çağkan Yapar, Fabian Jaensch, Jan~C. Hauffen, Francesco Pezone, Peter Jung,
    Saeid~K. Dehkordi, and Giuseppe Caire,
  \newblock ``{DEMUCS} for data-driven {RF} signal denoising,''
  \newblock in {\em Proc. IEEE Int. Conf. Acoust., Speech, Signal Process.
    Workshops (ICASSPW)}, Apr. 2024, pp. 95--96.
  
  \bibitem{defossez20}
  Alexandre Défossez, Gabriel Synnaeve, and Yossi Adi,
  \newblock ``Real time speech enhancement in the waveform domain,''
  \newblock in {\em Interspeech}, 2020, pp. 3291--3295.
  
  \bibitem{defossez21}
  Alexandre Défossez, Nicolas Usunier, Léon Bottou, and Francis Bach,
  \newblock ``Music source separation in the waveform domain,''
  \newblock in {\em Proc. Int. Conf. Learn. Represent. (ICLR)}, 2021,
  \newblock Available online: https://openreview.net/forum?id=HJx7uJStPH.
  
  \bibitem{IMEC2024}
  Mostafa Naseri, Jaron Fontaine, Ingrid Moerman, Eli De~Poorter, and Adnan
    Shahid,
  \newblock ``A {U-Net} architecture for time-frequency interference signal
    separation of {RF} waveforms,''
  \newblock in {\em Proc. IEEE Int. Conf. Acoust., Speech, Signal Process.
    Workshops (ICASSPW)}, Apr. 2024, pp. 91--92.
  
  \bibitem{hyvarinen1999fast}
  Aapo Hyvarinen,
  \newblock ``Fast and robust fixed-point algorithms for independent component
    analysis,''
  \newblock {\em {IEEE} Trans. Neural Netw.}, vol. 10, no. 3, pp. 626--634, 1999.
  
  \bibitem{belouchrani1997blind}
  Adel Belouchrani, Karim Abed-Meraim, J-F Cardoso, and Eric Moulines,
  \newblock ``A blind source separation technique using second-order
    statistics,''
  \newblock {\em IEEE Trans. Signal Process.}, vol. 45, no. 2, pp. 434--444,
    1997.
  
  \bibitem{yeredor2002non}
  Arie Yeredor,
  \newblock ``Non-orthogonal joint diagonalization in the least-squares sense
    with application in blind source separation,''
  \newblock {\em {IEEE} Trans. Signal Process.}, vol. 50, no. 7, pp. 1545--1553,
    2002.
  
  \bibitem{weiss2019maximum}
  Amir Weiss and Arie Yeredor,
  \newblock ``A maximum likelihood-based minimum mean square error separation and
    estimation of stationary {G}aussian sources from noisy mixtures,''
  \newblock {\em IEEE Trans. Signal Process.}, vol. 67, no. 19, pp. 5032--5045,
    July 2019.
  
  \bibitem{lancho2022data}
  Alejandro Lancho, Amir Weiss, Gary~C.F. Lee, Jennifer Tang, Yuheng Bu, Yury
    Polyanskiy, and Gregory~W. Wornell,
  \newblock ``Data-driven blind synchronization and interference rejection for
    digital communication signals,''
  \newblock in {\em IEEE Glob. Commun. Conf. (GLOBECOM)}, Dec. 2022, pp.
    2296--2302.
  
  \bibitem{bommasani2021opportunities}
  Rishi Bommasani, Drew~A Hudson, Ehsan Adeli, Russ Altman, Simran Arora, Sydney
    von Arx, Michael~S Bernstein, Jeannette Bohg, Antoine Bosselut, Emma
    Brunskill, et~al.,
  \newblock ``On the opportunities and risks of foundation models,''
  \newblock {\em arXiv:2108.07258}, 2021.
  
  \bibitem{fontaine2024}
  Jaron Fontaine, Adnan Shahid, and Eli~De Poorter,
  \newblock ``Towards a wireless physical-layer foundation model: {C}hallenges
    and strategies,''
  \newblock {\em arXiv:2403.12065}, Feb. 2024.
  
  \bibitem{alikhani24}
  Sadjad Alikhani, Gouranga Charan, and Ahmed Alkhateeb,
  \newblock ``Large wireless model ({LWM}): {A} foundation model for wireless
    channels,''
  \newblock {\em arXiv:2411.08872 [cs.IT]}, Nov. 2024.
  
  \end{thebibliography}

\end{document}